\documentclass[prd,twocolumn,eqsecnum,amsfonts,amssymb]{revtex4}

\usepackage{graphicx}
\usepackage{subfig}

\usepackage{bm}

\newcommand{\beq}{\begin{equation}}
\newcommand{\eeq}{\end{equation}}
\newcommand{\beqs}{\begin{eqnarray}}\newcommand{\eeqs}{\end{eqnarray}}

\begin{document}

\title{Exact Results for Average Cluster Numbers in Bond Percolation on
  Infinite-Length Lattice Strips} 

\author{Shu-Chiuan Chang$^a$ and Robert Shrock$^b$}

\affiliation{(a) \ Department of Physics, National Cheng Kung University,
  Tainan 70101, Taiwan} 

\affiliation{(b) \ C. N. Yang Institute for Theoretical Physics and
Department of Physics and Astronomy, \\
Stony Brook University, Stony Brook, New York 11794, USA }
\begin{abstract}

  We calculate exact analytic expressions for the average cluster
  numbers $\langle k \rangle_{\Lambda_s}$ on infinite-length strips
  $\Lambda_s$, with various widths, of several different lattices, as
  functions of the bond occupation probability, $p$.  It is proved
  that these expressions are rational functions of $p$.  As special
  cases of our results, we obtain exact values of $\langle k
  \rangle_{\Lambda_s}$ and derivatives of $\langle k
  \rangle_{\Lambda_s}$ with respect to $p$, evaluated at the critical
  percolation probabilities $p_{c,\Lambda}$ for the corresponding
  infinite two-dimensional lattices $\Lambda$.  We compare these exact
  results with an analytic finite-size correction formula and find
  excellent agreement.  We also analyze how unphysical poles in
  $\langle k \rangle_{\Lambda_s}$ determine the radii of convergence
  of series expansions for small $p$ and for $p$ near to unity. Our
  calculations are performed for infinite-length strips of the square,
  triangular, and honeycomb lattices with several types of transverse
  boundary conditions.
  
\end{abstract}

\maketitle


\section{Introduction}
\label{intro_section}

The study of percolation on lattice graphs elucidates the effect of
vacant sites and/or bonds on the connectedness properties of the
system. Here we shall consider bond percolation, in which the bonds of
the lattice are randomly present with probability $p$ and thus absent
with probability $1-p$.  Percolation is relevant for the analysis of
such phenomena as the flow of liquids through porous rock, electrical
conduction through composite materials, and the magnetic properties of
materials with lattice defects and impurities.  On an infinite lattice
$\Lambda$, as $p$ decreases from 1 to 0, the probability $P(p)$ for a
site to be part of an infinite connected cluster decreases and
vanishes at a critical value, $p_{c,\Lambda}$, remaining identically
zero for $0 \le p < p_{c,\Lambda}$.  Other quantities also behave
nonanalytically at $p=p_{c,\Lambda}$.  For example, as $p$ increases
toward $p_{c,\Lambda}$ from below, the average cluster size $S(p)$ diverges.
Thus, the percolation transition is a geometrical transition from a
region $0 \le p < p_{c,\Lambda}$ in which only finite connected
clusters exist, to a region $p_{c,\Lambda} \le p \le 1$ in which there
is a percolating cluster containing an infinite number of sites and
bonds.  The singularities in various quantities such as $P(p)$ and
$S(p)$ are described by a set of critical exponents depending only
on the dimensionality $d$ of $\Lambda$, but independent of the specific
type of lattice, and type (site or bond) of percolation (some reviews include
\cite{essam80}-\cite{perc_review2014}).

One of the interesting quantities in percolation is the average number
of (connected) clusters per site on a lattice graph $G$, in particular,
the limit as the number of sites $n \to \infty$,
\beq
\langle k \rangle_{ \{G \}} = \lim_{n \to \infty} n^{-1} \langle k
\rangle_G \ , 
\label{kdef}
\eeq
where $\{ G \}$ denotes the given $n \to \infty$ limit of the family
of $n$-vertex graphs $G$. Here, as in mathematical graph theory
\cite{bollobas}, a cluster is defined as a connecting subgraph of $G$,
including single sites. Since as $p \to 0$, there are no bonds, and
each site is a cluster, it follows that $\lim_{p \to 0} \langle k
\rangle_{ \{ G \} } = 1$.  On the other hand, as $p \to 1$, there is
just one cluster, namely $\Lambda$, so $\langle k \rangle_{ \{ G \}} =
0$.  This function $\langle k \rangle_{\{ G \}}$ is a monotonically
decreasing function of $p$ for $0 \le p \le 1$; it is continuous but
nonanalytic at $p=p_{c,\Lambda}$, with a finite singularity of the
form $(\langle k \rangle_{\{ G \}})_{sing.} \propto
|p-p_{c,\Lambda}|^{2-\alpha}$.  There is no exact solution for
$\langle k \rangle_\Lambda$ as a general function of $p$ for (site or
bond) percolation on a regular lattice of dimension $d \ge 2$,
although a solution has been calculated for the Bethe lattice
\cite{fe}.  Much has been learned from series expansions
\cite{essam80,es66,series,domb_pearce} and Monte Carlo simulations
\cite{newmanziff,perc_review2014}.

Although the critical exponents describing singularities in quantities
such as $P(p)$ and $S(p)$ are universal, the critical (threshold)
values of $p$ depend on the type (site or bond) of percolation and on
the type of lattice $\Lambda$.  For bond percolation on the
two-dimensional lattices considered here, exact expressions are known
for these critical percolation threshold values, $p_{c,\Lambda}$
\cite{exactpc}. The exact values of $\langle k \rangle_\Lambda$ on
each of these lattices $\Lambda$, evaluated at the respective critical
values $p=p_{c,\Lambda}$, have also been determined \cite{ziff97} (see
also \cite{tl,bta}), as have the finite-size corrections
\cite{kleban_ziff,zlk}.

In \cite{pc,pc2} we gave exact analytic calculations of average
cluster numbers $\langle k \rangle_{\Lambda_s}$ as functions of $p$ in
bond percolation for infinite-length strips, with various widths, of a
variety of lattices with certain transverse boundary conditions.
We also gave numerical values of $\langle k \rangle_{\Lambda_s}$ evaluated
at $p=p_{c,\Lambda}$ to five-digit accuracy.

In the present paper we report a far-reaching extension of this
earlier work, which has enabled us to substantially increase the
number of lattice strips for which we are able to calculate exact
analytical expressions for the average cluster numbers.  By
convention, the longitudinal (horizontal) direction along a given
lattice strip to taken to be the $x$ direction, and the transverse
(vertical) direction to be the $y$ direction.  We denote a given
infinite-length strip of a lattice $\Lambda$ with width $L_y$ sites
and with free (F) or periodic (P) transverse boundary conditions
(FBC$_y$, PBC$_y$) as $[\Lambda, (L_y)_F]$ or $[\Lambda,
  (L_y)_P]$. For the case of the square lattice, we have also
calculated $\langle k \rangle$ for the case of infinite-length strips
with self-dual (sd) transverse boundary conditions, and we denote
these strips as $[sq,(L_y)_{sd}]$. We will often use the compact
notation
\beq
\Lambda_s \equiv [\Lambda,(L_y)_{BC_y}] 
\label{lambdas}
\eeq
for infinite-length lattice strips (where the subscript $s$ stands for
``strip''). 

Our new results include the following:

\begin{enumerate}

\item
  A theorem showing that for an infinite-length lattice strip
  $[\Lambda,(L_y)_{BC_y}]$,
  the average number of clusters per site,
  $\langle k \rangle_{[\Lambda,(L_y)_{BC_y}]}$, is a rational
  function of the bond occupation probability, $p$.

\item
 Calculation of exact expressions for $\langle k
 \rangle_{[\Lambda,(L_y)_{BC_y}]}$ as functions of $p$, for a variety of
 infinite-length lattice strips with width $L_y$ and certain
 transverse boundary conditions. The lattices are square, triangular,
 and honeycomb. 

\item
  
 Calculation of the exact values of $\langle k
  \rangle_{[\Lambda,(L_y)_{BC_y}]}$ evaluated at the critical value of
  $p$ for the corresponding infinite 2D lattice, $\Lambda$,
  $p_{c,\Lambda}$, which we denote as
\beq
  \langle k \rangle_{[\Lambda,(L_y)_{BC_y}]}|_{p=p_{c,\Lambda}} \ . 
  \label{kscrit}
\eeq
The numerical values of these exact analytic expressions agree with
the numerical values that we presented in \cite{pc} for the respective
infinite-length strips.

\item

  A quantitative study of how these values
  approach the critical value $\langle k \rangle_{c,\Lambda}$ for the
  infinite two-dimensional lattice $\Lambda$ as the strip width $L_y$ 
  increases and, in particular, a comparison with the exact results
  from Ref. \cite{kleban_ziff,zlk} for the leading finite-size correction term
  in the case of periodic transverse boundary conditions: 
\beq  
\langle k \rangle_{[\Lambda,(L_y)_P]}|_{p=p_{c,\Lambda}} =
\langle k \rangle |_{c,\Lambda} + \frac{c_\Lambda \tilde b}{L_y^2}  + ... \ , 
\label{finite_size_correction}
\eeq
where we use the notation
\beq
\langle k \rangle_{c,\Lambda} \equiv
\langle k \rangle_\Lambda |_{p=p_{c,\Lambda}}  
\label{kcrit_lambda}
\eeq
for the average cluster number, per site, on the infinite two-dimensional
lattice $\Lambda$ evaluated at $p=p_{c,\Lambda}$, and the dots denote
higher-order terms in $1/L_y$.  The coefficient $\tilde b$
is \cite{kleban_ziff}
\beq
\tilde b = \frac{5\sqrt{3}}{24} = 0.360844 \ , 
\label{btilde}
\eeq
and $c_\Lambda$ is a mathematical constant that takes account of the
geometry of the lattice (see Eqs. (\ref{ctri}) and (\ref{chc})
below).  With this geometric relation incorporated, the
coefficient $\tilde b$ is universal.  Our exact results are in very
good agreement with this formula (\ref{finite_size_correction}),
including (i) the $(L_y)^{-2}$ dependence of the leading correction
term, (ii) the value of $\tilde b$, and (iii) the
universality with respect to lattice type.  The universality of $\tilde b$
was previously demonstrated from a comparative analysis of 
the square and triangular lattices in \cite{zlk}.  In this
context, we recall that our results for $\langle k
\rangle_{[\Lambda,(L_y)_{BC_y}]}$ and hence for the values $\langle k
\rangle_{[\Lambda,(L_y)_P]}|_{p=p_{c,\Lambda}}$, are independent of
the longitudinal boundary conditions imposed on the lattice strips.
For the comparison with $\langle
k\rangle_{\Lambda}|_{p=p_{c,\Lambda}}$ on the corresponding infinite
two-dimensional lattice $\Lambda$, we define the ratio
\beq
R_{[\Lambda,(L_y)_{BC_y}],c} = 
\frac{\langle k \rangle_{[\Lambda,(L_y)_{BC_y}]}|_{p=p_{c,\Lambda}}}
{\langle k \rangle_{c,\Lambda}} \ .
\label{kcrit_ratio}
\eeq

\item

  As a corollary of our theorem, a proof that for infinite-length
  square-lattice strips, the critical values $\langle k
  \rangle_{[sq,(L_y)_{BC_y}]}|_{p=p_{c,\Lambda}}$ are rational numbers
  and for infinite-length strips of the triangular and honeycomb
  lattices, they are rational functions of the quantity 
  $\sin(\pi/18)$ that appears in $p_{c,tri}$ and $p_{c,hc}$.

\item
  
Calculations of $d^j \langle k \rangle_{[\Lambda,(L_y)_{BC_y}]}/(dp)^j$ with
$j=1,2,3$, evaluated at $p=p_{c,\Lambda}$, for infinite-length lattice strips
$\Lambda_s$ with a resultant determination of coefficients in the expansion
of $\langle k \rangle_{[\Lambda,(L_y)_{BC_y}]}$ about this value
$p_{c,\Lambda}$.

\item 

 A study of the poles in $\langle k \rangle_{[\Lambda,(L_y)_{BC_y}]}$
 involving the determination of the pole or the complex-conjugate pair
 of poles closest to the origin in the complex $p$ plane, which thus
 set the radius of convergence of the small-$p$ series.  A
 corresponding analysis is given of the poles of $\langle k
 \rangle_{[\Lambda,(L_y)_{BC_y}]}$ in the complex $r$ plane, where
\beq
r \equiv 1-p \ . 
\label{r}
\eeq

\end{enumerate}

For reference, our results in \cite{pc} included analytic calculations of
$\langle k \rangle_{\Lambda_s}$ for $\Lambda_s=
[sq,2_F]$, $[sq,2_P]$, $[sq, 1_{sd}]$, $[tri,2_F]$, $[tri,2_P]$,
and $[hc,2_P]$.  We also presented numerical calculations of
$\langle k \rangle_{\Lambda_s}$ and
$\langle k \rangle_{\Lambda_s}|_{p=p_{c,\Lambda}}$
for several other infinite-length strips.  We review these here in comparison
with our new results.

In statistical mechanics, it had been very valuable to use
high-temperature series expansions of thermodynamic quantities to
determine the critical temperature. This was determined via the
estimate of the radius of convergence of these series. However, the
application of this procedure in studies of percolation encountered a
complication, namely that the radii of convergence of these series
expansions were typically determined not by the actual critical values
of $p_{c,\Lambda}$ or $r_{c,\Lambda}$, but instead by unphysical
singularities in the respective complex $p$ plane and $r$ plane that
lie closer to the origin than $p_{c,\Lambda}$ or $r_{c,\Lambda}$
\cite{series,domb_pearce}.  Although this complication was
circumvented, e.g., by the use of Pad\'e approximants, to get accurate
determinations of critical behavior at the percolation transition, it
raises an intriguing question, namely whether one would encounter the
presence of similar unphysical singularities in analyses of exact
expressions for average cluster numbers on infinite-length lattice
strips.  Our work in \cite{pc} provided some initial insight into this
question.  Our present results go substantially further in answering
this question, since we have now succeeded in calculating exact
expressions for $\langle k \rangle_{\Lambda_s}$ for considerably
greater strip widths.  Indeed, one of the interesting results of our
study of the poles in the exact expressions for $\langle k
\rangle_{\Lambda_s}$ on various infinite-length lattice strips
$\Lambda_s$ is that we find that, as the strip width $L_y$ increases,
it is generic that there is a pole on the negative real axis or a
complex-conjugate pair of poles in the complex $p$ plane closer to the
origin than the value $p_{c,\Lambda}$ for the infinite lattice, and
similarly for the pole(s) in the $r$-plane.


\section{Background} 
\label{background_section}

In this section we review some relevant background. We begin with an
important connection between percolation and the Potts model. For the
sake of generality, let us consider the $q$-state Potts model on a
connected graph $G=(V,E)$ defined by its set of sites (vertices) $V$
and its set of bonds (called ``edges'' in mathematical graph terminology),
$E$. In graph theory, a percolation cluster is a connected subgraph of
$G$.  The partition function of the $q$-state Potts model on $G$ is
\cite{fk}
\beq
Z(G,q,v) = \sum_{G' \subseteq G} q^{k(G')} v^{e(G')} \ , 
\label{pottscluster}
\eeq
where $G'=(V,E')$ is a spanning subgraph of $G$, i.e., a subgraph
containing all of the sites in $G$ and a subset $E' \subseteq E$ of
the bonds of $G$, and $e(G')$ is the number of bonds in $G'$.
In the thermal context, $v=e^K-1$ is a
temperature-dependent Boltzmann variable, with $K=J/(k_BT)$, where
$J$ is the spin-spin coupling in the Potts Hamiltonian,
${\cal H}=-J\sum_{e_{ij}} \delta_{\sigma_i,\sigma_j}$, where
$e_{ij}$ is the bond connecting sites $i$ and $j$ in $G$.  The
dimensionless free energy is then defined as
\beq
f(\{ G \},q,v) = \lim_{n \to \infty} \frac{1}{n} \ln[Z(G,q,v)] \ , 
\label{f}
\eeq
where, as above, $n$ denotes the number of sites in $G$ and 
$\{ G \}$ denotes the $N \to \infty$ limit of $G$.
Now in $f(\{ G \}, q,v)$, set $v=v_p$, where
\beq
v_p = \frac{p}{1-p} \ .
\label{vp}
\eeq
Then the average number of clusters per site is 
\beq
\langle k \rangle_{\{ G \}} = \frac{\partial f(\{ G \},q,v_p)}
                               {\partial q} \bigg |_{q=1}
\ .
\label{k}
\eeq
Now we specialize to the case where $G$ is a lattice graph.  The relation
(\ref{k}) leads to the inference that the percolation transition
on these lattices is in the universality class of the $q$-state
Potts model in the limit where $q \to 1$, with critical exponents
$\alpha=-2/3$, $\beta=5/36$, $\gamma=43/18$, etc. \cite{crit,wurev,qeq1}
and an associated
conformal field theory having a Virasoro algebra with central charge $c=0$
\cite{cardy92}
for the case of dimensionality $d=2$ relevant here. 

In \cite{pc,pc2} we used the relation (\ref{k}) together with 
our earlier exact calculations of $f$ on strips of
lattices with arbitrarily great length and fixed width, with various
transverse boundary bounditions (BC$_y$) to obtain new analytic
expressions and numerical values for $\langle k \rangle_{\Lambda_s}$ for
infinite-length strips of these types.  

For a lattice $\Lambda$, in the thermodynamic limit, the average
cluster number per site has the following expansion in the local
neighborhood of $p_{c,\Lambda}$ 
\beqs
\langle k \rangle_{\Lambda} &=& \langle k \rangle_{c,\Lambda} +
a_{1,\Lambda_s}(p-p_{c,\lambda}) + a_{2,\Lambda_s}(p-p_{c,\lambda})^2 
\cr\cr
&+& {\cal A}_{\Lambda,\pm}|p-p_{c,\lambda}|^{2-\alpha} \ , 
\label{kcrit_expansion}
\eeqs
where $\alpha=-2/3$ for $d=2$, as noted above, and the amplitudes
${\cal A}_{\Lambda,\pm}$ refer to the limits $p-p_{c,\Lambda} \to
0^\pm$, respectively. Thus, $\langle k \rangle_{\Lambda}$ has a finite
branch-point singularity at $p=p_{c,\Lambda}$. A recent discussion of
the coefficients in this expansion is \cite{ziff2017} (where $\langle
k \rangle_{\Lambda}$ is defined per bond rather than per site).

A theorem that we present below shows that on an infinite-length strip
of a lattice $\Lambda$ with width $L_y$ and some prescribed transverse
boundary conditions $BC_y$, $\langle k \rangle_{\Lambda_s}$, evaluated
at $p=p_{c,\Lambda}$, is a rational function of $p$. Although it is
therefore meromorphic, none of its poles occur in the physical
interval $p \in [0,1]$.  Hence, for $p$ in this interval, it has a
Taylor series expansion, and if one evaluates this at the value of $p$
equal to the critical value for the infinite lattice, $p_{c,\Lambda}$,
then one obtains
\beq
\langle k \rangle_{\Lambda_s}|_{p=p_{c,\Lambda}} =
  \langle k \rangle_{\Lambda_s,c} + \sum_{j=1}^\infty 
  a_{j,\Lambda_s} (p-p_{c,\Lambda})^j \ , 
\label{kstrip_expansion}
\eeq
where
\beq
a_{j,\Lambda_s} = \frac{1}{j!} \frac{d^j \langle k \rangle_{\Lambda_s}}
{(dp)^j}\Big |_{p=p_{c,\Lambda}} \ . 
\label{adef}
\eeq
As our results in \cite{pc} showed, and our current results further
demonstrate, for a given infinite-length strip $[\Lambda,(L_y)_{BC_y}]$, as
$L_y$ increases, $\langle k \rangle_{\Lambda_s}|_{p=p_{c,\Lambda}}$
  approaches the critical value $\langle k \rangle_{c,\Lambda}$ for the
  infinite two-dimensional lattice.

The known values of critical bond occupation
probabilities for the square (sq), triangular (tri), and honeycomb (hc)
lattices are \cite{exactpc} (see also \cite{tl,bta}) 
\beq
p_{c,sq} = \frac{1}{2} \ , 
\label{pc_sq}
\eeq
\beq
p_{c,tri} = 2 \sin \bigg ( \frac{\pi}{18} \bigg ) = 0.347296 \ , 
\label{pc_tri}
\eeq
and
\beq
p_{c,hc} = 1 - p_{c,tri} = 1 - 2 \sin \bigg ( \frac{\pi}{18} \bigg )
= 0.652704 \ . 
\label{pc_hc}
\eeq
Here and below, floating-point values are given to the indicated number
of significant figures. It will be convenient to introduce the shorthand 
symbol
\beq
s \equiv \sin \Big ( \frac{\pi}{18} \Big ) \ . 
\label{s}
\eeq
Exact analytic expressions for 
$\langle k \rangle_{c,\Lambda}$ were presented in \cite{ziff97} (see also
related results in \cite{tl,bta}): 
\beq
\langle k \rangle_{c,sq} = \frac{3\sqrt{3}-5}{2} = 0.0980762 \ , 
\label{kcrit_sq}
\eeq
\beqs
\langle k \rangle_{c,tri} &=& \frac{35}{4}-\frac{3}{p_{c,tri}} =
\frac{-6+35s}{4s} \cr\cr
&=& 0.111844 \ , 
\label{kcrit_tri}
\eeqs
and
\beqs
\langle k \rangle_{c,hc} &=& \frac{1}{2}(\langle k \rangle_{c,tri}
+ p_{c,tri}^3 ) = \frac{-6+31s+24s^2}{8s} \cr\cr
&=& 0.0768667 \ . 
\label{kcrit_hc}
\eeqs

On a lattice $\Lambda$ with coordination number $\Delta_\Lambda$, the
small-$p$ series expansion for $\langle k \rangle_\Lambda$ has the
generic form $\langle k \rangle_\Lambda = 1 - (\Delta_\Lambda/2)p +
...$, where the dots indicate higher-order terms.  For the square,
triangular, and honeycomb lattices, the small-$p$ series expansions
are \cite{series}
\beq
\langle k \rangle_{sq} = 1 - 2p + p^4 + 2p^6 - 2p^7 + 7p^8 + O(p^9) \ , 
\label{ksq_pseries}
\eeq
\beqs
\langle k \rangle_{tri} &=& 1 - 3p + 2p^3 + 3p^4 + 3p^5 + 3p^6 + 6p^7 + O(p^9) 
\ , \cr\cr
&&
\label{ktri_pseries}
\eeqs
and
\beq
\langle k \rangle_{hc} = 1 - \frac{3}{2}p + \frac{1}{2}p^6 + \frac{3}{2}p^{10}
+ O(p^{11}) \ . 
\label{khc_pseries}
\eeq
These have been calculated to higher order than shown here, but we will only
need the expansions to these respective orders for comparison with the
small-$p$ expansions of our exact expressions for average cluster numbers on
infinite-length strips of various lattices with specified transverse
boundary conditions. 

It has also been valuable to calculate Taylor series
expansions of average cluster numbers in terms of the expansion variable $r$
for small $r$.  On (the thermodynamic limit of) a lattice $\Lambda$ with 
coordination number $\Delta_\Lambda$,
the small-$r$ series expansion for $\langle k \rangle_{\Lambda}$ has the
generic form $\langle k \rangle_{\Lambda} = r^{\Delta_\Lambda} + ...$ where the
$...$ indicate higher-order terms.  For the square, triangular, and honeycomb
lattices, the small-$r$ series expansions are
\cite{series} 
\beq
\langle k \rangle_{sq} = r^4 + 2r^6 - 2r^7 + 7r^8 + O(r^9) \ , 
\label{sq_rseries}
\eeq
\beq
\langle k \rangle_{tri} = r^6 + 3r^{10} - 3r^{11} + 2r^{12} + O(r^{14}) \ , 
\label{tri_rseries}
\eeq
and
\beq
\langle k \rangle_{hc} = r^3 + \frac{3}{2}r^4 + \frac{3}{2}r^6 + O(r^7) \ . 
\label{hc_rseries}
\eeq
%


\section{Calculational Methods}
\label{methods_section}

We consider strip graphs of a lattice $\Lambda$ of finite width $L_y$
and arbitrarily great length $m=L_x$, with a given set of longitudinal
and transverse boundary conditions. For these strip graphs, the Potts
model partition function $Z$ has the form of a finite sum of $m$'th powers:
\beq
Z([\Lambda, L_x,L_y,BC_x,BC_y],q,v) = \sum_j \kappa_j (\lambda_j)^m  \ , 
\label{zform}
\eeq
where $\kappa_j$ are coefficients and $\lambda_j$ are
certain functions that depend the type of strip, but are independent of
the length, $m$.  In the limit of infinite length, $m \to \infty$, this
sum is dominated by the $\lambda$ of largest magnitude, so that the 
reduced dimensionless free energy is 
\beq
f([\Lambda,(L_y)_{BC_y}],q,v) = \frac{1}{L_y}\ln[\lambda_{dom,[\Lambda,
      (L_y)_{BC_y}]}] \ . 
\label{fform}
\eeq
In previous work, we have determined the $\lambda$ functions, and in
particular, $\lambda_{dom}$, for a number of lattice strips $\Lambda_s$ (e.g.,
\cite{a}-\cite{sdg}. As was shown in this earlier work, the above-mentioned
dominant $\lambda$ function, and hence the resultant reduced free energy $f$,
are independent of the type of longitudinal boundary conditions used for the
finite-$m$ lattice strips.  We shall make use of a general property of
$Z(G,q,v)$, which holds for any graph $G$, namely
\beq
Z(G,q=1,v)=(v+1)^{e(G)} \ , 
\label{zgq1}
\eeq
where, as above, $e(G)$ denotes the number of edges (bonds) on $G$. This
follows because if $q=1$, then the Potts model Hamiltonian ${\cal H}$
reduces simply to ${\cal H} = -Je(G)$, so
\beq
Z(G,q=1,v)=e^{Ke(G)}=(v+1)^{e(G)} \ . 
\label{zq1}
\eeq
For a $\Delta$-regular graph $G$, $e(G)=(\Delta/2)n(G)$. More generally, for 
a graph which is not $\Delta$-regular, 
one can define an effective vertex degree $\Delta_{\rm{eff}}$ 
(e.g., \cite{wn}) as
\beq
\Delta_{\rm{eff}} = \lim_{n(G) \to \infty} \frac{2e(G)}{n(G)} \ . 
\label{delta_eff}
\eeq
Hence, for a family of $\Delta$-regular lattice strip graphs $\Lambda_s$,
Eq. (\ref{fform}) applies for $q=1$ with 
\beq
\lambda_{dom,\Lambda_s}|_{q=1} = (v+1)^{(\Delta/2)L_y} \ ,
\label{lamdom_q1}
\eeq
and similarly for non-$\Delta$-regular graphs, with $\Delta$ replaced
by $\Delta_{{\rm eff}}$.  In particular, for the application to percolation,
setting $v=v_p=p/(1-p)$, we have
\beq
\lambda_{dom,\Lambda_s}|_{q=1,v=v_p} =
\Big ( \frac{1}{1-p} \Big )^{(\Delta/2)L_y} \ . 
\label{lamdom_q1_perc}
\eeq

Each of the $\lambda$ functions appearing in Eq. (\ref{zform}), and,
in particular, the dominant $\lambda$, is a
solution to an algebraic equation,
\beq
\sum_{j=0}^{j_{max}} \kappa_{\Lambda_s,j} \, (\lambda_{\Lambda_s})^j = 0 \ , 
\label{lameq}
\eeq
where the coefficients $\kappa_{\Lambda_s,j}$ are polynomials in $q$ and $v$.
For many strip graphs, $j_{max}$ in the equation of the form 
(\ref{lameq}) for the dominant $\lambda$ is $j_{max} \ge 5$,
so that one cannot solve for $\lambda_{dom,\Lambda_s}$
in terms of radicals.  Fortunately,
however, one does not need to do this; all that one needs to do is to
calculate $\lambda_{dom,\Lambda_s}$ and $d\lambda_{dom,\Lambda_s}/dq$, both
evaluated at $q=1$, for insertion into Eq. (\ref{k}). We can do this as
follows. 
Differentiating Eq. (\ref{lameq}) with respect to $q$ and solving for
$\lambda_{dom,\Lambda_s}$, we have
\beq
\frac{d\lambda_{dom,\Lambda_s}}{dq} = 
-\frac{\sum_{j=0}^{j_{max}}(\lambda_{dom,\Lambda_s})^j \, 
\frac{d\kappa_{\Lambda_s,j}}{dq} }
      {\sum_{j=1}^{j_{max}} j \, \kappa_{\Lambda_s,j} \,
       (\lambda_{dom,\Lambda_s})^{j-1}} \ . 
\label{dlamdq}
\eeq
Evaluating this equation at $q=1$ and $v=v_p$, we have
\beq
\frac{d\lambda_{dom,\Lambda_s}}{dq} \Big |_{q=1,v=v_p} = 
-\frac{\sum_{j=0}^{j_{max}}(1-p)^{-j} \, 
[\frac{d\kappa_{\Lambda_s,j}}{dq}]|_{q=1,v=v_p}}
  {\sum_{j=1}^{j_{max}} j \, [\kappa_{\Lambda_s,j}|_{q=1,v=v_p}] \,
       (1-p)^{1-j}} \ . 
\label{dlamdq_q1}
\eeq
This is a powerful result, because it means that in calculating $\langle k
\rangle_{\Lambda_s}$, one does not have to actually solve for the dominant root
$\lambda_{dom,\Lambda_s}$, but instead, only use its derivative evaluated 
at $v=v_p$ and $q=1$, which can be expressed as a rational function of $p$.

Having explained our method of calculation, we next discuss the
analytic structure of the results and their pertinence to series
expansions.  Because the Potts model is a discrete spin model, the
series expansions for $\langle k \rangle_{\Lambda_s}$ for small $p$ or
for small $r$ are Taylor series expansions, with finite radii of
convergence. Owing to the fact that $v_p=p/(1-p)$, a small-$p$
expansion for a (bond or site) percolation problem is formally
analogous to a high-temperature expansion of the corresponding Potts
model.  Normally, a high-temperature expansion in a Potts model has a
radius of convergence equal to the critical point.  However, the radii
of convergence of Taylor series expansions around both $p=0$ and $p=1$
were typically set by unphysical singularities, and these radii of
convergence were less than the distance from the expansion point to
the physical singularity, $p_{c,\Lambda}$, for the small-$p$
expansions and $r_{c\Lambda} = 1-p_{c, \Lambda}$ for small-$r$
expansions \cite{series,domb_pearce}.  We showed in \cite{pc} using
the exact expressions that we calculated for $\langle k
\rangle_{\Lambda_s}$ on infinite-length, finite-width lattice strips,
that these expressions also exhibited poles nearer to the origin in
the complex $p$ plane than the respective value of $p_{c,\Lambda}$ on
the infinite two-dimensional lattice.  Similarly, we showed that these
expressions, as functions of $r$, exhibited poles closer to the origin
in the complex $r$ plane than $r_{c,\Lambda}=1-p_{c,\Lambda}$ for the
corresponding infinite two-dimensional lattices. Thus, the
calculations of $\langle k \rangle_{\Lambda_s}$ on infinite-length
lattice strips $\Lambda_s$ in \cite{pc} provided insight into the
influence of unphysical poles in the small-$p$ and small-$r$ series
expansions on infinite two-dimensional lattices. Our new results
provide further insight into this phenomenon.

Our results on radii of convergence and pole structure are based on a
general property that we have proved above, that $\langle k
\rangle_{\Lambda_s}$ is a rational function of $p$ and hence also of
$r=1-p$.  For a given infinite-length strip $\Lambda_s$ of the lattice
$\Lambda$ of finite width $L_y$ and specified transverse boundary
conditions $BC_y$, let us denote the set of poles in the complex $p$
plane as $p_{\Lambda_s,i}$ with the index $i$ enumerating the number
of poles. For each infinite-length lattice strip $\Lambda_s$, we
determine the pole or complex-conjugate pair of poles closest to the
origin, which thus determines the radius of convergence of the
small-$p$ series.  In a similar way, our exact expressions $\langle k
\rangle_{\Lambda_s}$ as functions of $r$ provide insight into this,
since we can determine the poles in each of them and, in particular,
the pole or complex-conjugate pair of poles closest to the origin in
the complex $r$ plane, which thus set the radius of convergence of the
respective small-$r$ series expansions of $\langle k
\rangle_{\Lambda_s}$. It should be noted that it is not the case that
there is a simple relation between the pole(s) nearest to the origin
in the $p$ plane and the pole(s) nearest to the origin in the complex
$r$ plane.  To illustrate this, let us consider a hypothetical
example, in which, for an infinite-length lattice strip $\Lambda_s$,
the exact expression for the average cluster number, $\langle k
\rangle_{\Lambda_s}$, has poles at $p=-0.4$ and $p=0.7$.  The pole
nearest to the origin in the $p$ plane is at $p=-0.4$, so the radius
of convergence of the small-$p$ series expansion of $\langle k
\rangle_{\Lambda_s}$ is 0.4.  In this hypothetical example, the poles
in $\langle k \rangle_{\Lambda_s}$, expressed as a function of $r$,
are at $r=0.3$ and $r=1.4$, so the radius of convergence of the
small-$r$ series is 0.3.  Thus, although there is a 1-1
correspondence between the full set of poles of $\langle k \rangle$ in
the complex $p$ and $r$ planes, it is not, in general, true that the
nearest pole to the origin in the complex $r$ plane, is equal to 1
minus the value of the nearest pole to the origin in the complex $r$
plane.

A word is in order concerning how the longitudinal and transverse
directions of our lattice strips relate to the lattice vectors.  For
the square-lattice strips, we take these longitudinal and transverse
directions to be the lattice axes. The strips of the triangular
lattice are constructed by starting with a square-lattice strip with
the same boundary conditions and adding diagonal bonds to each square,
say from the lower left site to the upper right site of each square.
A picture of several illustrative finite-length sections of these
triangular-lattice strips was included as Fig. 1 in
Ref. \cite{t}. Pictures of finite-length sections of strips of the
honeycomb (brick) lattice were given as Figs. 16 and 18 in
Ref. \cite{hca}.  In Refs. \cite{dg,sdg} we presented results for
square-lattice strip graphs with several types of self-dual transverse
boundary conditions (see also \cite{p}).  These all yield the same
expression for $\langle k \rangle_{sq,(L_y)_{sd}}$.  To construct a
strip of the square lattice with one type of self-dual boundary
condition, one starts with a square-lattice strip of length $L_x$ and
width $L_y$ vertices and periodic longitudinal boundary conditions and
adds bonds connecting each site on the upper side of the strip to a
single external vertex.  For a picture of a finite-length section of
this self-dual square-lattice strip graph, we refer the reader to
Fig. 1 of Ref. \cite{sdg}. A second type of self-dual square-lattice
strip is constructed by starting with this first type of self-dual
strip and then adding bonds connecting each site on the lower side of
the strip to a second external vertex.

The expressions for the effective coordination numbers 
$\Delta_{{\rm eff}}$, as defined in Eq. (\ref{delta_eff}), 
for the infinite-length strips that we consider here are listed below: 
\beq
\Delta_{[sq,(L_y)_F],{\rm eff}} = 4 - \frac{2}{L_y} \ ,
\label{delta_eff_sqf}
\eeq
\beq
\Delta_{[tri,(L_y)_F],\rm{eff}} = 6 - \frac{4}{L_y} \ ,
\label{delta_eff_trif}
\eeq
and
\beq
\Delta_{[hc,(L_y)_F],\rm{eff}} = 3 - \frac{1}{L_y} \ . 
\label{delta_eff_hcf}
\eeq
For the infinite-length limit of the second type of self-dual square-lattice
strip, we have
\beq
\Delta_{[sq,(L_y)_{sd}],\rm{eff}} = 4 \ . 
\label{delta_eff_sqsd}
\eeq
%


\section{Some General Properties}
\label{general_properties_section}

In this section we prove several general theorems and discuss some general
structural features of our exact calculations of
average cluster numbers $\langle k \rangle_{\Lambda,BC_y}$ for infinite-length
strips of lattices $\Lambda$ with finite width $L_y$ and various transverse
boundary conditions $BC_y$. (As noted before, all results are independent of
the longitudinal boundary conditions used for a given lattice strip.)


\subsection{$\langle k \rangle_{[\Lambda,(L_y)_{BC_y}]}$ is a Rational Function
of $p$}

We first prove an important theorem stating that for 
an infinite-length strip graph $\Lambda_s = [\Lambda,(L_y)_{BC_y}]$, 
the average cluster number per site, $\langle k \rangle_{\Lambda_s}$, 
is a rational function of $p$ and hence also of $r$. That is, 
\beq
\langle k \rangle_{\Lambda_s} =\frac{N_{\Lambda_s}}{D_{\Lambda_s}} \ , 
\label{kform}
\eeq
where, $N$ and $D$ denote numerator and denominator polynomials in $p$. 
In factorized form, 
\beq
\langle k \rangle_{\Lambda_s} = 
\frac{\prod_{i=1}^{{\rm deg}_p(N_{\Lambda_s})} (1-p/a_i)}
     {\prod_{j=1}^{{\rm deg}_p(D_{\Lambda_s})} (1-p/b_j)} \ .
\label{kformfac}
\eeq
This applies to an arbitrary two-dimensional lattice, and is not limited to
the specific types of lattices (square, triangular, and honeycomb) for which we
calculate $\langle k \rangle_{[\Lambda,(L_y)_{BC_y}]}$ here. 
To prove this theorem, we note that, from Eq. (\ref{k}), 
\beq
\langle k \rangle_{\Lambda_s} = \frac{1}{L_y} \, \frac{
\Big (\frac{d\lambda_{dom,\Lambda_s}}{dq}\Big ) \Big |_{q=1,v=v_p}}
{\lambda_{dom,\Lambda_s}|_{q=1,v=v_p}} \ . 
\eeq
From Eqs. (\ref{lamdom_q1}) and (\ref{dlamdq_q1}), it follows that 
this is a rational function of $p$. 

This is a very interesting and useful result, because naively, if one were to
make direct use of $\langle k \rangle_{\Lambda_s}$ via Eq. (\ref{k}) as the
derivative of $f=\ln(\lambda_{dom,\Lambda_s})$ with respect to $q$, evaluated
at $q=1$, one might naturally think that it would be necessary first to
calculate $\lambda_{dom,\Lambda_s}$. With strips for which this is possible, 
the algebraic equation that yields $\lambda_{dom,\Lambda_s}$ is of degree 2
to 4, so $\lambda_{dom,\Lambda_s}$ would be an algebraic, but not rational,
function of $q$, and for wider strips, the algebraic equation that yields
$\lambda_{dom,\Lambda_s}$ is of degree 5 or higher, so one would not be able 
to solve for $\lambda_{dom,\Lambda_s}$ analytically at all. As our method of 
calculation presented in Section \ref{methods_section} shows, one can avoid
this problem by making use of Eq. (\ref{dlamdq_q1}), which does not require
solving for $\lambda_{dom.}$ itself as a general function of $q$, but only
the evaluation at $v=v_p$ and $q=1$. 

From our theorem in Eq. (\ref{kform}) above, it follows that 
$\langle k \rangle_{[\Lambda,(L_y)_{BC_y}]}$ is a meromorphic function of $p$,
with poles at
\beqs
p &=& p_{[\Lambda,(L_y)_{BC_y}],j} = b_j \ , \quad j=1,...,
{\rm deg}_p(D_{[\Lambda,(L_y)_{BC_y}]}) \ . \cr\cr
&& 
\label{kppoles}
\eeqs
Clearly, when expressed as a function of $r$,
$\langle k \rangle_{\Lambda,(L_y)_{BC_y}}$ is again a rational function 
\beq
\langle k \rangle_{[\Lambda,(L_y)_{BC_y}]} =
\frac{N_{[\Lambda,(L_y)_{BC_y}],r}}
     {D_{[\Lambda,(L_y)_{BC_y}],r}} \ , 
\label{kformr}
\eeq
where $N_{[\Lambda,(L_y)_{BC_y}],r}$ and $D_{[\Lambda,(L_y)_{BC_y}],r}$ are
polynomials in $r$ of degree
${\rm deg}_r(N_{[\Lambda,(L_y)_{BC_y}],r})$ and 
${\rm deg}_r(D_{[\Lambda,(L_y)_{BC_y}],r})$, respectively, with
\beq
    {\rm deg}_p(N_{[\Lambda,(L_y)_{BC_y}]})=
    {\rm deg}_r(N_{[\Lambda,(L_y)_{BC_y}],r})
\label{degp_eq_degr_numerator}
\eeq
and
\beq
    {\rm deg}_p(D_{[\Lambda,(L_y)_{BC_y}]})=
    {\rm deg}_r(D_{[\Lambda,(L_y)_{BC_y}],r})
\ . 
\label{degp_eq_degr_denominator}
\eeq
Furthermore, there is a 1-1 correspondence between the poles of
$\langle k \rangle_{[\Lambda,(L_y)_{BC_y}]}$ in the $p$ plane and in the
$r$ plane.


\subsection{$\langle k \rangle_{[sq,(L_y)_{BC_y}]}|_{p=p_{c,sq}}$ is a Rational 
Number}

An important corollary of our theorem in Eq. (\ref{kform}) is that in 
the case of square-lattice strips, when one evaluates 
$\langle k \rangle_{[sq,(L_y)_{BC_y}]}$ at $p=p_{c,sq}=1/2$, 
the result, namely, $\langle k \rangle_{[sq,(L_y)_{BC_y}]}|_{p=p_{c,sq}}$, 
is a rational number.  

Although this property does not hold for strips of other lattices such as
triangle or honeycomb, one has an analogous result, namely that because
$p_{c,tr}$ is a polynomial of the quantity $s \equiv \sin(\pi/18)$
defined in Eq. (\ref{s}) and $p_{c,hc}$ is a rational function of $s$, 
$\langle k \rangle_{[tri,(L_y)_{BC_y}]}|_{p=p_{c,tri}}$  and
$\langle k \rangle_{[hc, (L_y)_{BC_y}]}|_{p=p_{c,hc}}$ are
rational functions of $s$. 


\subsection{Agreement with Universal Finite-Size Scaling Formula}

As noted in the introduction, our exact results for
$\langle k \rangle_{[\Lambda,(L_y)_{BC_y}]}$ evaluated at $p=p_{c,\Lambda}$
enable us to make several comparisons, to check agreement with (a) the
values $\langle k \rangle_{c,\Lambda}$ and (b) with the formula
(\ref{finite_size_correction}) from \cite{kleban_ziff,zlk} for the finite-size
correction term, involving three individual checks: (i) the $(L_y)^{-2}$
dependence on strip width of the leading finite-size correction, (ii)
the coefficient $\tilde b$ in Eq. (\ref{btilde}), and (iii) the universality
with respect to lattice type. 
For the comparison (b), we define a constant
\beq
\tilde b_{[\Lambda,(L_y)_{BC_y}]} = c_\Lambda^{-1} L_y^2 \Big [
  \langle k \rangle_{[\Lambda,(L_y)_P]} -
  \langle k \rangle_{c,\Lambda} \Big ] \ . 
\label{bform}
\eeq
The $c_\Lambda^{-1}$ in Eq. (\ref{finite_size_correction}) is a
geometrical factor connected with the relation between the area $A_p$
of a regular $p$-sided polygon and the length $a$ of a side (= lattice
spacing in our case), $A_p =pa^2/[4\tan(\pi/p)]$. The role of
$c_{tri}$ in the universality of $\tilde b$ for the square and
triangular lattices was shown in \cite{zlk}. For the lattices that we
consider, $c_{sq}=1$ and, with our notational conventions in
\cite{t,ta} and \cite{hca},
\beq
c_{tri} = \frac{\sqrt{3}}{2} 
\label{ctri}
\eeq
and
\beq
c_{hc} = \frac{1}{\sqrt{3}} \ , 
\label{chc}
\eeq
so that $c_{tri}\tilde b = 5/16$ and $c_{hc}\tilde b = 5/24$.
Agreement with the formula (\ref{finite_size_correction}) requires
that, as the width, $L_y$, of the infinite-length strip increases, the
quantity $\tilde b_{[\Lambda,(L_y)_P]}$ should approach the value
$\tilde b = 5\sqrt{3}/24$, independent of lattice type.  We find
excellent agreement with both (a) and all three parts (i)-(iii) of
property (b).  Our results are listed in Table \ref{b_table} and show
excellent concordance, in particular, with part (iii) of condition
(b), for all of the types of lattice that we consider, namely, square,
triangular, and honeycomb.  Quantitatively, as is
evident in Table \ref{b_table}, the ratios $\tilde b_{[sq,5_P]}/\tilde
b$, $\tilde b_{[hc,4_P]}/\tilde b$, and $\tilde b_{[tri,4_P]}/\tilde
b$ differ from unity by the respective amounts $1 \times 10^{-2}$, $4
\times 10^{-3}$, and $1 \times 10^{-4}$. These ratios are thus quite
close to unity even for these modest-width strips.


\subsection{Property of Poles for Square-Lattice Strips with Periodic and
Self-Dual Transverse Boundary Conditions}

We find an interesting special property of the expressions for
$\langle k \rangle_{[sq,(L_y)_P]}$ and $\langle k
\rangle_{[sq,(L_y)_{sd}]}$, i.e., of the average cluster numbers for
the infinite-length strips of the square lattice with width $L_y$ and
either periodic or self-dual (sd) transverse boundary conditions.  For
each such strip, we find that the denominator of $\langle k
\rangle_{[sq,(L_y)_P]}$ (resp. $\langle k \rangle_{[sq,(L_y)_{sd}]}$),
expressed as a function of $p$, is the same as this denominator
expressed as a function of $r$, with the interchange $r
\leftrightarrow p$. That is, for the strips with periodic transverse
boundary conditions, given
\beq
\langle k \rangle_{[sq,(L_y)_P]} = \frac{N_{[sq,(L_y)_P]}}
  {D_{[sq,(L_y)_P]}} \ , 
\label{ksqLyp_general}
\eeq
with
\beq
N_{[sq,(L_y)_P]} =
(1-p)^m ( L_y + \sum_{\ell} c_{[sq,(L_y)_P],\ell} \, p^\ell  )  \ , 
\label{ksqLyp_general_num}
\eeq
where $m$ is a certain power depending on $L_y$, and
\beq
D_{[sq,(L_y)_P]} =
L_y (1+ \sum_{\ell} d_{[sq,(L_y)_P],\ell} \, p^\ell )   \ , 
\label{ksqLyp_general_denp}
\eeq
the denominator polynomial has the form
\beq
D_{[sq,(L_y)_P]} = L_y (1+ \sum_{\ell} d_{[sq,(L_y)_P],\ell} \, r^\ell ) \ . 
\label{ksqLyp_general_denr}
\eeq
The same property expressed in Eqs. (\ref{ksqLyp_general})-
(\ref{ksqLyp_general_denr}) also holds for the square-lattice strips
with self-dual boundary conditions.  Hence, the set of poles of
$\langle k \rangle_{[sq,(L_y)_P]}$ in the $p$ plane have the same values
as the set of poles in the $r$ plane, and similarly for the set of
poles of $\langle k \rangle_{[sq,(L_y)_{sd}]}$. Note that this equality
of coefficients for $p^j$ and $r^j$ terms in
Eqs. (\ref{ksqLyp_general_denp}) and (\ref{ksqLyp_general_denr}) is
not implied by the fact that that the denominator of a given strip,
expressed in terms of $p$, is equal to this denominator, written in
terms of $r=1-p$.  Indeed, the special coefficient equality embodied
in Eqs. (\ref{ksqLyp_general_denp}) and (\ref{ksqLyp_general_denr}) is
not true for the other infinite-length, finite-width strips for which
we have obtained exact calculations of the average cluster number.


\subsection{Some Properties of the Derivatives
  $\frac{d^j\langle k \rangle_{\Lambda_s}}{(dp)^j}$}

We have found several properties of the $j$'th derivatives
$d^j\langle k \rangle_{\Lambda_s}/(dp)^j$ for general infinite-length
lattice strips.  First, as a corollary of our theorem (\ref{kform}) that
$\langle k \rangle_{\Lambda_s}$ is a rational function of $p$, it follows
that the $j$'th derivative
$d^j\langle k \rangle_{\Lambda_s}/(dp)^j$ is also a rational function of $p$
and that any evaluation of this function for rational $p$ is a rational
number.

Second, for an infinite-length strip graph $\Lambda_s$ which is
$\Delta$-regular, 
\beq
\frac{d\langle k \rangle_{\Lambda_s}}{dp}\Big |_{p=0} = -\frac{\Delta}{2} \ . 
\label{dkdp_p0}
\eeq
If the infinite-length strip graph $\Lambda_s$ is not $\Delta$-regular, then
this relation holds with $\Delta$ replaced by $\Delta_{\rm{eff}}$ on the
right-hand side.

Third, for all infinite-length lattice strips $\Lambda_s$ with
$\Delta \ge 3$ (in the $\Delta$-regular case) or, more generally,
$\Delta_{\rm{eff}} \ge 3$, 
\beq
\frac{d\langle k \rangle_{\Lambda_s}}{dp}\Big |_{p=1} = 0 \ . 
\label{dkdp_p1}
\eeq
This property (\ref{dkdp_p1}) holds for all of the lattice strips
considered here, given our condition on the vertex degree.  (This
condition excludes the 1D strip, for which $\Delta=2$ and $\langle
k \rangle_{1D}=1-p$, so $d\langle k \rangle_{1D}/dp=-1$ independent of
$p$.)


\subsection{Structural Properties of $\frac{d^j\langle k 
\rangle_{[sq,(L_y)_{P,sd}]}}{(dp)^j}$ and $a_{j,[sq,(L_y)_{BC_y}]}$}

For infinite-length strips of the square lattice with width $L_y$ and
either periodic or self-dual boundary transverse conditions, we find
several general results concerning $\frac{d^j\langle k
  \rangle_{[sq,(L_y)_{P,sd}]}}{(dp)^j}$ and $a_{j,sq,(L_y)_{BC_y}}$
for $1 \le j \le 3$. For compact notation, we will denote
infinite-length square-lattice strips with either of these two types
of transverse boundary conditions as $[sq,(L_y)_{P,sd}]$. 
First, $d^3 \langle k
\rangle_{[sq,(L_y)_{P,sd}]}/(dp)^3$ has the symmetry property that under
a replacement of $p \to 1-p$, this third derivative reverses in sign:
\beq
 \frac{d^3\langle k \rangle_{[sq,(L_y)_{P,sd}]}}{(dp)^3}(p) =
-\frac{d^3\langle k \rangle_{[sq,(L_y)_{P,sd}]}}{(dp)^3}(1-p) \ , 
\label{d3kdp_signreversal}
\eeq
where the $(p)$ and $(1-p)$ indicate the arguments of the respective
functions.  Consistent with this symmetry property, we find that
\beq
\frac{d^3\langle k \rangle_{[sq,(L_y)_{P,sd}]}}{(dp)^3} \quad {\rm
  contains \ the \ factor} \ (1-2p) \ . 
\label{d3kdp_factors}
\eeq

Concerning evaluations of $\langle k \rangle_{[sq,(L_y)_{P,sd}]}$ at
the critical value of $p$ for the infinite lattice, namely $p_{c,sq}=1/2$,
which yield the coefficients $a_{1,[sq,(L_y)_{P,sd}]}$, we find that 
\beq
a_{1,[sq,(L_y)_{P,sd}]} = -1 \ .
\label{a1_sqLpsd}
\eeq
This agrees with Ref. \cite{ziff2017}, when one takes account of the
fact that we define $\langle k \rangle$ per site here, while
Ref. \cite{ziff2017} defines $\langle k \rangle$ per bond. Our
calculations of $a_{1,[sq,(L_y)_F]}$ for the strips with free
transverse boundary conditions) are consistent with the inference that
these coefficients approach the value $-1$ in the $L_y \to \infty$
limit. The fact that the value is already reached for finite $L_y$ on
the square-lattice strips with periodic or self-dual transverse
boundary conditions shows the advantage in the use of these latter
boundary conditions, since they remove boundary effects and render the
strip graphs 4-regular.  Finally, given that $d^3\langle k
\rangle_{[sq,(L_y)_{P,sd}]}/(dp)^3$ contains the factor $(2p-1)$, it
follows that
\beq
a_{3,[sq,(L_y)_{P,sd}]} = 0 \ .
\label{a3_sqLpsd}
\eeq
%


\subsection{Relation Between Small-$p$ and Small-$r$ Series Expansions
  of $\langle k \rangle_{[sq,(L_y)_{P,sd}]}$ }

From our calculations of $\langle k \rangle_{[sq,(L_y)_P]}$ and $\langle
k \rangle_{[sq,(L_y)_{sd}]}$, we find that in all cases, the small-$p$
and small-$r$ Taylor series expansions of $\langle k
\rangle_{[sq,(L_y)_P]}$, and, separately, the small-$p$ and small-$r$
Taylor series expansions of $\langle k \rangle_{[sq,(L_y)_{sd}]}$, are
closely related and are of the form
\beq
\langle k \rangle_{[sq,(L_y)_{P,sd}]} = 1-2p + \sum_{\ell=L_y}^\infty
h_{[sq,(L_y)_{P,sd}],\ell} \, p^\ell
\label{ksqLypsd_pseries_form}
\eeq
and
\beq
\langle k \rangle_{[sq,(L_y)_{P,sd}]} = \sum_{\ell=L_y}^\infty
h_{[sq,(L_y)_{P,sd}],\ell} \, r^\ell \ , 
\label{ksqLypsd_rseries_form}
\eeq
where, as before, the subscript $P,sd$ means that the equality holds
separately for the square-lattice strips with periodic or self-dual
transverse boundary conditions.  Thus, except for the first two terms
in the small-$p$ series, all of the coefficients in both of these
series, from the respective $O(p^{L_y})$ and $O(r^{L_y})$ orders to
infinity, are the same.  Since the radii of convergence of these
series are determined by the behavior of the small-$p$ and small-$r$
series as the order goes to infinity (e.g., by the ratio test), this
equality of the coefficients is in accord with the property discussed
in the previous subsection, that the poles are at the same positions
in the $p$ plane and in the $r$ plane for each of these strips, so
that the pole (or complex-conjugate pair of poles) that is closest to
the origin is the same in the $p$ and $r$ planes, and hence the
small-$p$ and small-$r$ series expansions have the same radius of
convergence.  In contrast, for other infinite-length, finite-width
strips of various lattices, the radius of convergence of the small-$p$
expansion is not, in general, equal to the radius of convergence of
the small-$r$ expansion.


\subsection{Some General Properties of the 
Numerator and Denominator Polynomials 
in $\langle k \rangle_{\Lambda_s}$ }

For many of the infinite-length, finite-width lattice strips $\Lambda_s$ 
for which we have calculated the exact
expressions $\langle k \rangle_{\Lambda_s}$, we find that the
degree of the numerator, as a polynomial in $p$ or $r$ is greater, by
one unit, than the degree of the denominator, i.e.,
\beq
{\rm deg}_p(N_{\Lambda_s})={\rm deg}_p(D_{\Lambda_s})+1 
\label{degree_numden_relation}
\eeq
for these strips.  These include the $[sq,(L_y)_{BC_y}]$ strips with
$BC_y=F, \ P, \ sd$; the $[hc,(L_y)_{BC_y}]$ strips with $BC_y=F, \ P$,
and the $tri,(L_y)_F$ strips.  
This is not the case with the $[tri,(L_y)_P]$ 
strips. For the $[tri,(L_y)_P]$ strips for which we have obtained
$\langle k \rangle_{[tri,(L_y)_P]}$, namely those with widths $L_y=2,3,4$,
we find that
\beq
{\rm deg}_p(N_{[tri,(L_y)_P}])={\rm deg}_p(D_{[tri,(L_y)_P}])+2L_y \ . 
\label{degree_numden_rel_trip}
\eeq
Calculations of $\langle k \rangle_{\Lambda_s}$ for larger values of $L_y$ 
would be necessary to determine if these patterns persist for wider strips. 

We find that the numerator $N_{[\Lambda,(L_y)_{BC_y}]}$ in
$\langle k \rangle_{[\Lambda,(L_y)_{BC_y}]}$ always contains
a prefactor (abbreviated PF) equal to $(1-p)=r$ raised to a certain
power depending on $[\Lambda,(L_y)_{BC_y}]$, which we denote as
${\rm deg}[PF(N_{[\Lambda,(L_y)_{BC_y}]})]$.  This power is equal to the
minimum power of $r$ in the small-$r$ expansion of
$N_{[\Lambda,(L_y)_{BC_y}]}$.  In Table \ref{kform_table}
we list the values of ${\rm deg}(N_{[\Lambda,(L_y)_{BC_y}]})$, ${\rm
  deg}[PF(N_{[\Lambda,(L_y)_{BC_y}]})]$ and ${\rm
deg}(D_{[\Lambda,(L_y)_{BC_y}]})$, for the strips for which we have
calculated the average cluster numbers $\langle k
\rangle_{[\Lambda,(L_y)_{BC_y}]}$.


\section{Strips of the Square Lattice}
\label{sq_section}


\subsection{Square-Lattice Strips with $(L_y)_{BC_y}=1_F, \ 2_F$} 
\label{sq1f_section}

We recall that for the one-dimensional line, $L_y=1$, an elementary
calculation yields $\langle k \rangle_{1D} = 1-p$.  In Ref. \cite{pc}
we calculated $\langle k \rangle_{[sq,2_F]}$. We review this result
here for comparison with our new results:
\beqs
\langle k \rangle_{[sq,2_F]} &=& \frac{(1-p)^2(2+p-2p^2)}{2(1-p^2+p^3)}
\cr\cr\cr
&=& \frac{r^2(1+3r-2r^2)}{2(1-r+2r^2-r^3)} \ .  \quad \quad 
\label{ksq2f}
\eeqs
As noted above, in Table \ref{kform_table} we list the degrees of the 
numerator and denominator of $\langle k \rangle_{[sq,2_F]}$ as polynomials in 
$p$ or equivalently in $r$, together with the degree of the prefactor 
$(1-p)^2$. At $p=p_{c,sq}$, $\langle k \rangle_{[sq,2_F]}$ has the value
\beq
\langle k \rangle_{[sq,2_F]}|_{p=p_{c,sq}} = \frac{2}{7} =
0.285714 \ . 
\label{ksq2f_pc}
\eeq
In Table \ref{kcrit_table} we list this critical value.  It is of
interest to compare the critical value (\ref{ksq2f_pc}) with $\langle
k \rangle_{c,sq}$ on the infinite square lattice.  For this purpose,
we list the values of the ratio (\ref{kcrit_ratio}) for the present
lattice strips and others in Table \ref{kcrit_table}.  Tables
\ref{kform_table} and \ref{kcrit_table} also list the corresponding
results for the other infinite-length, finite-width lattice strips
with various widths and transverse boundary conditions denoted $BC_y$
for which we have calculated $\langle k
\rangle_{[\Lambda,(L_y)_{BC_y}]}$. 

It is instructive to study derivatives of $\langle k \rangle_{[sq,2_F]}$
and to apply these to calculate the coefficients $a_{[sq,2_F],j}$ in
Eq. (\ref{kstrip_expansion}) for the first several values of $j$.
Doing this, we obtain the results
\beq
a_{1,[sq,2_F]} = -\frac{59}{7^2} = -1.204082
\label{a1_sq2f_pc}
\eeq
\beq
a_{2,[sq,2_F]} = \frac{316}{7^3} = 0.921283
\label{a2_sq2f_pc}
\eeq
and
\beq
a_{3,[sq,2_F]} = \frac{2872}{7^4} = 1.196168 \ . 
\label{a3_sq2f_pc}
\eeq
We list these values in Table \ref{aj_table}, which also lists the
analogous values of these coefficients for other infinite-length
lattice strips.  In Eqs. (\ref{a1_sq2f_pc})-(\ref{a3_sq2f_pc}) we have
indicated the factorizations of the denominators. In general, the
numerators of these expressions do not have similarly simple
factorizations; for example, the numerators of $a_{[sq,2_F],j}$ for
$j=1,2,3$ have the respective factorizations 59, $2^2 \cdot 79$, and
$2^3 \cdot 359$.  For an infinite two-dimensional lattices $\Lambda$,
the leading singularity in $\langle k \rangle_{\Lambda}$ occurs in the
$|p-p_{c,\Lambda}|^{2-\alpha} = |p-p_{c,\Lambda}|^{8/3}$ term in
Eq. (\ref{kcrit_expansion}), but, as a consequence of our theorem
(\ref{kform}), it follows that $\langle k \rangle_{\Lambda_s}$ does
not have any branch-point singularities such as
$|p-p_{c,\Lambda}|^{8/3}$.

The first few terms of the small-$p$ and small-$r$ Taylor series
expansions of $\langle k \rangle_{[sq,2_F]}$ are listed in Table
\ref{series_table}.  Of the three poles in $\langle k
\rangle_{[sq,2_F]}$ in the complex $p$ plane, the \underline{n}earest
\underline{p}ole (denoted with subscript $np$) to the origin is at
\beq
p_{[sq,2_F],np} = -0.754878 \ . 
\label{ppole_sq2f}
\eeq
The magnitude of this pole is equal to the radius of convergence of the
small-$p$ series expansion of $\langle k \rangle_{[sq,2_F]}$. In the
complex-$r$ plane, there are two poles in $\langle k
\rangle_{[sq,2_F]}$ nearest to the origin, namely the
complex-conjugate pair
\beq
r_{[sq,2_F],np} = 0.122561 \pm 0.744862i \ , 
\label{rpole_sq2f}
\eeq
with magnitude $|r_{[sq,2_F],np}|=|p_{sq,2_f,np}|=0.754878$. We list
these values of the poles of $\langle k \rangle_{[sq,2_F]}$ nearest to
the origin in the $p$ and $r$ plane, and their magnitudes, in Table
\ref{pole_table}.  This table also lists the corresponding poles and
magnitudes for the other infinite-length lattice strips with various
widths and transverse boundary conditions that we consider in this
paper.  This table includes information concerning whether the the
resulting radii of convergence of the small-$p$ and small-$r$ series
expansions of $\langle k \rangle_{[\Lambda,(L_y)_{BC_y}]}$ are smaller
or larger than the corresponding critical values $p_{c,\Lambda}$ and
$r_{c,\Lambda}=1-p_{c,\Lambda}$ for these other lattice strips. .


\subsection{$3_F$ Square-Lattice Strip}
\label{sq3f_section}

As the first of our new results for explicit expressions of
$\langle k \rangle_{\Lambda_s}$ on infinite-length lattice strips, we
present our calculation of this average cluster number for the
$[sq,3_F]$ strip: 
\begin{widetext}
\beqs
\langle k \rangle_{[sq,3_F]} &=&
\frac{(1-p)^3(3+4p-3p^2-8p^3+9p^4+12p^5-26p^6+9p^7+11p^8-11p^9+3p^{10})}
{3(1+p-p^2)(1-p+p^2)(1-p^2-2p^3+6p^4-2p^5-3p^6+3p^7-p^8 )} 
\cr\cr\cr
&=&
\frac{r^3(3+4r-7r^2+5r^3+8r^4-31r^5-61r^7+47r^8-19r^9+3r^{10} )}
{3(1+r-r^2)(1-r+r^2)(1-r-r^2+9r^3-14r^4+13r^5-10r^6+5r^7-r^8 )}  \ . 
\cr\cr
&&
\label{ksq3f}
\eeqs
\end{widetext}
As indicated, it is useful to express this and other average cluster
numbers $\langle k \rangle_{[\Lambda,(L_y)_{BC_y}]}$ as functions of $p$
and also, equivalently, as functions of $r$.
When evaluated at $p=p_{c,sq}$, $\langle k \rangle_{[sq,3_F]}$ has the value
\beq
\langle k \rangle_{[sq,3_F]}|_{p=p_{c,sq}} = \frac{147}{670} = 0.219403 \ . 
\label{ksq3f_pc}
\eeq

The first three coefficients $a_{[sq,3_F],j}$ with $j=1,2,3$, are 
\beq
a_{1,[sq,3_F]} = -\frac{16355}{3 \cdot (67)^2}= -1.214450
\label{a1_sq3f_pc}
\eeq
\beq
a_{2,[sq,3_F]} = \frac{297238112}{3^3 \cdot 5^2 \cdot (67)^3} = 1.464119
\label{a2_sq3f_pc}
\eeq
and
\beq
a_{3,[sq,3_F]} = \frac{1004115424}{3^3 \cdot (67)^4} = 1.845528 \ . 
\label{a3_sq3f_pc}
\eeq

The Taylor series expansions of $\langle k \rangle_{[sq,3_F]}$ for small
$p$ and $r$ are
\beqs
\langle k \rangle_{[sq,3_F]} &=& 1-\frac{5}{3}p+\frac{2}{3}p^4+p^6-p^7+ 
\frac{7}{3}p^8-4p^9+O(p^{10}) \cr\cr
&&
\label{ksq3f_pseries}
\eeqs
and 
\beqs
\langle k \rangle_{[sq,3_F]} &=& r^3+\frac{7}{3}r^4+2r^5-\frac{11}{3}r^6-9r^7
 -\frac{49}{3}r^8 \cr\cr
&+&\frac{86}{3}r^9+O(r^{10}) \ . 
\label{ksq3f_rseries}
\eeqs
For comparison with results for other strips, we list the first few
terms of these series in Table \ref{series_table}.

Of the 12 poles of $\langle k \rangle_{[sq,3_F]}$ in the complex $p$ plane, the 
ones nearest to the origin are a complex-conjugate pair at
\beq
p_{[sq,3_F],np} = -0.400758 \pm 0.399068i \ , 
\label{ppole_sqf3_np}
\eeq
of magnitude
\beq
|p_{[sq,3_F],np}| = 0.565564 \ , 
\label{ppole_sqf3_np_mag}
\eeq
which is therefore the radius of convergence of the small-$p$ series
for $\langle k \rangle_{[sq,3_F]}$, as indicated in Table \ref{pole_table}. 
In the complex-$r$ plane, the pole of
$\langle k \rangle_{[sq,3_F]}$ nearest to the origin occurs at the value
\beq
r_{[sq,3_F],np} = -0.411578 \ , 
\label{rpole_sqf3_np}
\eeq
of magnitude $|r_{[sq,3_F],np}| = 0.411578$, which is thus the radius
of convergence of the small-$r$ series for $\langle k
\rangle_{[sq,3_F]}$.  Note that for this infinite-length strip, we
have the generic behavior that $1-|r_{[sq,3_F],np}|$ is not equal to
$|p_{[sq,3_F],np}|$.


\subsection{$4_F$ Square-Lattice Strip}
\label{sq_4f_section}

We calculate 
\beq
\langle k \rangle_{[sq,4_F]} = \frac{N_{[sq,4_F]}}{D_{[sq,4_F]}} \ , 
\label{ksq4f}
\eeq
where the numerator and denominator polynomials $N_{[sq,4_F]}$ and
$D_{[sq,4_F]}$ are given in Eqs. (\ref{ksq4f_numerator}) and
(\ref{ksq4f_denominator}) in the appendix. 
At $p=p_{c,sq}$, $\langle k \rangle_{[sq,4_F]}$ has the value
\beq
\langle k \rangle_{[sq,4_F]}|_{p=p_{c,sq}}=\frac{27229}{145196}=0.187533 \
. 
\label{ksq4f_pc}
\eeq
This and the other exact values of $\langle k
\rangle_{\Lambda,(L_y)_{BC_y}}$ evaluated at $p=p_{c,\Lambda}$ for
moderately wide strips of the square lattice with various transverse
boundary conditions do not have particularly simple
factorizations. For example, the factorizations of the numerator and
denominator of Eq. (\ref{ksq4f_pc}) are $27229 = 73 \cdot 373$ and
$145196=2^2 \cdot 36299$.

For the coefficient $a_{1,[sq,4_F]}$ we calculate
\beq
a_{1,[sq,4_F]} = -\frac{14241087916}{11858556609} = -1.200912 \ . 
\label{a1_sq4f_pc}
\eeq
The analytic expressions for the coefficients
$a_{2,[sq,4_F]}$ and $a_{3,[sq,4_F]}$ are sufficiently lengthy that we only 
give the floating-point values:
\beq
a_{2,[sq,4_F]} = 1.833688
\label{a2_sq4f_pc}
\eeq
and
\beq
a_{3,[sq,4_F]} = 2.2777505 \ . 
\label{a3_sq4f_pc}
\eeq

The Taylor series expansions of $\langle k \rangle_{[sq,4_F]}$ for small
$p$ and $r$ are listed in Table \ref{series_table}. 
Of the 45 poles of $\langle k \rangle_{[sq,4_F]}$ in the complex $p$ plane, 
the pole nearest to the origin is
\beq
p_{[sq,4_F],np} = -0.492588 \ , 
\label{ppole_sqf4_np}
\eeq
which thus sets the radius of convergence of the small-$p$ series for
$\langle k \rangle_{[sq,4_F]}$ as $|p_{[sq,3_F],np}| = 0.492588$. Note
that this is smaller than the physical singularity of
$\langle k \rangle_{sq}$ on the infinite square lattice, at
$p=p_{c,sq}=1/2$, as indicated in Table \ref{pole_table}.  In the
complex-$r$ plane, the poles of $\langle k \rangle_{sq,4_F}$ nearest
to the origin are the complex-conjugate pair
\beq
r_{[sq,4_F],np} = -0.317578 \pm 0.244625i \ , 
\label{rpole_sqf4_np}
\eeq
of magnitude $|r_{[sq,4_F],np}| = 0.400871$, which is thus the radius
of convergence of the small-$r$ series expansion of $\langle k
\rangle_{[sq,4_F]}$. Note that for this strip, we again have
the generic behavior that $1-|r_{[sq,4_F],np}|$ is not equal to
$|p_{[sq,4_F],np}|$.


\subsection{$2_P$ Square-Lattice Strip}
\label{sq2p_section}

Results for infinite-length strips with periodic transverse boundary
conditions have the advantage, relative to those with free transverse
boundary conditions, that they are free of boundary effects, although,
of course, they still reflect the finite transverse size of the strips.
The expression for $\langle k \rangle_{[sq,2_P]}$ is \cite{pc}
\beqs
\langle k \rangle_{[sq,2_P]} &=& \frac{(1-p)^2(2-3p^2+2p^3)}
        {2(1+p-p^2)(1-p+p^2)} \cr\cr\cr
        &=& \frac{r^2(1+3r^2-2r^3)}{2(1+r-r^2)(1-r+r^2)} \ . 
\label{ksq2p}
\eeqs
The property that the denominator of $\langle k \rangle_{[sq,2_P]}$,
expressed as a function of $p$ and as a function of $r$, has the same
form, with the interchange of variables $p \leftrightarrow r$, is a
general consequence of the structural property noted above in Eqs.
(\ref{ksqLyp_general_num})-(\ref{ksqLyp_general_denr}).  At
$p=p_{c,sq}=1/2$, the average cluster number $\langle k
\rangle_{[sq,2_P]}$ has the value
\beq
\langle k \rangle_{[sq,2_P]}\Big|_{p=p_{c,sq}} = \frac{1}{5}  \ . 
\label{ksq2p_pc}
\eeq

To illustrate the general structural properties noted above in Eqs.
(\ref{d3kdp_signreversal}) and (\ref{d3kdp_factors}), we exhibit the
functional forms of the derivatives $(1/j!)d^j\langle k
\rangle_{[sq,2_P]}/(dp)^j$ for $1 \le j \le 3$.  These are
\begin{widetext}
\beq
\frac{d\langle k \rangle_{[sq,2_P]}}{dp} =
-\frac{(1-p)(2-p)(1+p-p^2-p^3+p^4-p^5+p^6)}
{[(1+p-p^2)(1-p+p^2)]^2} \ , 
\label{dkdp_sq2p}
\eeq
\beq
\frac{1}{2}\frac{d^2\langle k \rangle_{[sq,2_P]}}{(dp)^2} =
\frac{1+21p^2-34p^3+3p^4+28p^6-24p^7+6p^8}
{2[(1+p-p^2)(1-p+p^2)]^3} \ , 
\label{d2kdp2_sq2p}
\eeq
and
\beq
\frac{1}{3!}\frac{d^3\langle k \rangle_{[sq,2_P]}}{(dp)^3} =
\frac{2p(1-p)(1-2p)(4+2p+7p^2-16p^3+4p^4+2p^5+4p^6-4p^7+p^8)}
{[(1+p-p^2)(1-p+p^2)]^4} \ . 
\label{d3kdp3_sq2p}
\eeq
\end{widetext}
When evaluated at $p=p_{c,sq}=1/2$, $d\langle k\rangle_{[sq,2_P]}/dp = -1$
and $d^3\langle k \rangle_{[sq,2_P]}/(dp)^2=0$, in accord with our general
findings indicated in Eqs. (\ref{a1_sqLpsd}) and (\ref{a3_sqLpsd}). 
For $a_{[sq,2_P],2}$ we calculate
\beq
a_{2,[sq,2_P]} = \frac{112}{75} = \frac{2^4 \cdot 7}{3 \cdot 5^2} = 1.493333 
\ . 
\label{a2_sq2p}
\eeq

The pole of $\langle k \rangle_{[sq,2_P]}$ nearest to the origin in the $p$ 
plane occurs at $p_{[sq,2_P],np}=-0.618034$, which sets the radius of the 
small-$p$ series expansion for $\langle k \rangle_{sq,2_P}$. In accord with the
general structure property noted in Eqs.
(\ref{ksqLyp_general})-(\ref{ksqLyp_general_denr}). 
the pole of $\langle k \rangle_{[sq,2_P]}$ nearest to the origin in the $r$ 
plane occurs at the same value, $r_{[sq,2_P],np}=-0.618034$, which 
sets the radius of convergence of the small-$r$ series expansion of 
$\langle k \rangle_{[sq,2_P]}$. 


\subsection{$3_P$ Square-Lattice Strip}
\label{sq3p_section}

Here, for $L_y=3$, we calculate
\begin{widetext}
\beqs
\langle k \rangle_{[sq,3_P]} &=& \frac{(1-p)^3(3+3p-3p^2-14p^3+18p^4-p^5-13p^6
+11p^7-3p^8)}{3(1-p^2-2p^3+11p^4-11p^5-p^6+10p^7-10p^8+5p^9-p^{10})}\cr\cr\cr
&=& \frac{r^3(1+3r+5r^2-5r^3-7r^4+16r^5-20r^6+13r^7-3r^8)}
{3(1-r^2-2r^3+11r^4-11r^5-r^6+10r^7-10r^8+5r^9-r^{10})} \ . 
\label{ksq3p}
\eeqs
\end{widetext}
At $p=p_{c,sq}=1/2$, this has the value
\beq
\langle k \rangle_{[sq,3_P]}\Big|_{p=p_{c,sq}} = \frac{11}{78} =
0.1410256 \ . 
\label{ksq3p_pc}
\eeq

As is evident from Eqs. (\ref{ksq2p}) and (\ref{ksq3p}), and is also
true for all of the other $(L_y)_P$ strips of the square lattice for
which we have calculated $\langle k \rangle_{[sq,(L_y)_P]}$, the poles
in the complex $p$ and $r$ planes have the same values.  The
coefficients $a_{1,[sq,3_P]}$ and $a_{3,[sq,3_P]}$ are given by our
general results (\ref{a1_sqLpsd}) and (\ref{a3_sqLpsd}). For
$a_{2,[sq,3_P]}$ we calculate
\beq
a_{2,[sq,3_P]} = \frac{77024}{34983} = 2.201755
\label{a2_sq3p}
\eeq

The first few terms of the small-$p$ and small-$r$ Taylor series
expansions of $\langle k \rangle_{[sq,3_P]}$ are given in Table
\ref{series_table}.  Since the poles of $\langle k \rangle_{[sq,3_P]}$
are the same when expressed in the variables $p$ and $r$, it follows
that the poles nearest to the origin in the comples $p$ and $r$ planes
have the same value. This is
\beq
p_{[sq,3_P],np} = r_{[sq,3_P],np} = -0.354731 \pm 0.319907i \ ,
\label{ppole_sq3p_np}
\eeq
with magnitude 
\beq
|p_{[sq,3_P],np}| = |r_{[sq,3_P],np}|=0.477676 \ . 
\label{ppole_sq3p_np_abs}
\eeq
These poles thus determine the radii of convergence of the respective small-$p$
and small-$r$ Taylor series expansions of $\langle k \rangle_{[sq,3_P]}$ 
as 0.477676.  As indicated in Table \ref{pole_table},
this radius of convergence is smaller than $p_{c,sq}=r_{c,sq}=1/2$.


\subsection{$4_P$ Square-Lattice Strip}
\label{sq4p_section}

We calculate
\beq
\langle k \rangle_{[sq,4_P]} = \frac{N_{[sq,4_P]}}{D_{[sq,4_P]}} \ ,
\label{ksq4p}
\eeq       
where the numerator and denominator polynomials $N_{[sq,4_P]}$ and
$D_{[sq,4_P]}$,
which are rather lengthy, are given in Eqs. (\ref{ksq4p_numerator}) and
(\ref{ksq4p_denominator}) in the appendix.
At $p=p_{c,sq}=1/2$, $\langle k \rangle_{[sq,4_P]}$ has the value
\beq
\langle k \rangle_{[sq,4_P]}\Big|_{p=p_{c,sq}} = \frac{677}{5572} =
0.121500 \ . 
\label{ksq4p_pc}
\eeq
For $a_{2,[sq,4_P]}$ we calculate
\beq
a_{2,[sq,4_P]} = \frac{3398556656}{1298160381} = 2.617979 \ . 
\label{a2_sq4p}
\eeq

In $\langle k \rangle_{[sq,4_P]}$, the nearest poles
to the origin in both the complex $p$ plane and the complex $r$ plane are at
\beq
p_{[sq,4_P],np} = r_{[sq,4_P],np} = -0.424294 \ , 
\label{ppole_sq4p_np}
\eeq
which determine the radii of convergence of the respective
small-$p$ and small-$fr$ Taylor series expansions of 
$\langle k \rangle_{[sq,4_P]}$. This radius of convergence is again 
smaller than $p_{c,sq}=r_{c,sq}=1/2$.


\subsection{$5_P$ Square-Lattice Strip}
\label{sq5p_section}

We calculate 
\beq
\langle k \rangle_{[sq,5_P]} = \frac{N_{[sq,5_P]}}{D_{[sq,5_P]}} \ ,
\label{ksq5p}
\eeq       
where $N_{[sq,5_P]}$ and $D_{[sq,5_P]}$ are given in Eqs.
(\ref{ksq5p_numerator}) and (\ref{ksq5p_denominator}) in the appendix. 
At $p=p_{c,sq}=1/2$, this has the value
\beq
\langle k \rangle_{[sq,5_P]} |_{p=p_{c,sq}} = \frac{85013}{753370} =
0.112844 \ .
\label{ksq5p_pc}
\eeq
This is only 15 \% larger than the value for the infinite square lattice:
\beq
R_{[sq,5_P],c} = 1.150571 \ , 
\label{ksq5p_crit_ratio}
\eeq
where $R_{[\Lambda,(L_y)_{BC_y}],c}$ was defined in Eq. (\ref{kcrit_ratio}). 

For $a_{2,[sq,5_P]}$ we calculate
\beq
a_{2,[sq,5_P]} = \frac{1275302677055206848}{439932074289972983} = 2.898863 \ . 
\label{a2_sq5p}
\eeq

The small-$p$ and small-$r$ Taylor series expansions of $\langle k
\rangle_{[sq,5_P]}$ are given in Table \ref{series_table}.  Of the 62
poles of $\langle k \rangle_{[sq,5_P]}$ when expressed as a function
of $p$, which are the same when expressed as a function of $r$, the
nearest poles to the origin in both the complex $p$ and $r$ planes are
the complex-conjugate pair
\beq
p_{[sq,5_P],np} = r_{[sq,5_P],np} = -0.371844 \pm 0.169863i 
\label{ppole_sq5p_np}
\eeq
with magnitude
\beq
|p_{[sq,5_P],np}| = |r_{[sq,5_P],np}| = 0.408805 \ . 
\label{ppole_sq5p_np_abs}
\eeq
These poles thus determine the radii of convergence of the respective
small-$p$ and small-$r$ Taylor series expansions of
$\langle k \rangle_{[sq,5_P]}$ as 0.408805. As was the case
with the $3_P$ and $4_P$ strips of the square lattice, this radius of
convergence is smaller than $p_{c,sq}=r_{c,sq}=1/2$. 


\subsection{Square-Lattice Strips with Self-Dual Transverse Boundary Conditions
and $L_y=1$}
\label{sqd1_section}

Since the square lattice is self-dual, it is also useful to employ
boundary conditions for strip graphs of the square lattice that obey
this property even for finite $L_x$ and $L_y$ \cite{p,dg,sdg}.  We
denote the average cluster number for the strip of the square lattice
with width $L_y$ and self-dual transverse boundary conditions as
$\langle k \rangle_{[sq,(L_y)_{sd}]}$. 

In \cite{pc} we
calculated $\langle k \rangle_{[sq,1_{sd}]}$, which is
\beq
\langle k \rangle_{[sq,1_{sd}]} = \frac{(1-p)^3}{1-p+p^2} =
\frac{r^3}{1-r+r^2} \ . 
\label{ksq1sd}
\eeq
At $p=p_{c,sq}=1/2$, this has the value
\beq
\langle k \rangle_{[sq,1_{sd}]}|_{p=p_{c,sq}} = \frac{1}{6} = 0.166667 \ .
\label{ksq1sd_pc}
\eeq

As a particularly simple illustration of the
general structural properties indicated in 
Eqs.~ (\ref{d3kdp_signreversal}) and (\ref{d3kdp_factors}), we exhibit the
explicit expressions for $(1/j!)d^j\langle k
\rangle_{[sq,1_{sd}]}/(dp)^j$ for $1 \le j \le 3$.  These are
\beq
\frac{d\langle k \rangle_{[sq,1_{sd}]}}{dp} =
-\frac{(1-p)^2(2+p^2)}{(1-p+p^2)^2} \ , 
\label{dkdp_sq1sd}
\eeq
\beq
\frac{1}{2}\frac{d^2\langle k \rangle_{[sq,1_{sd}]}}{(dp)^2} =
\frac{3p(1-p)}{(1-p+p^2)^3} \ , 
\label{d2kdp2_sq1sd}
\eeq
and
\beq
\frac{1}{3!}\frac{d^3\langle k \rangle_{[sq,1_{sd}]}}{(dp)^3} =
\frac{(1-2p)(1+2p-2p^2)}{(1-p+p^2)^4} \ . 
\label{d3kdp3_sq1sd}
\eeq

The coefficients $a_{1,[sq,1_{sd}]}$ and $a_{3,[sq,1_{sd}]}$ are given by our
general results (\ref{a1_sqLpsd}) and (\ref{a3_sqLpsd}). For
$a_{2,[sq,1_{sd}]}$ we calculate
\beq
a_{2,[sq,1_{sd}]} = \frac{16}{9} = 1.777778 \ . 
\label{a2_sq1sd}
\eeq

The poles of $\langle k \rangle_{[sq,1_{sd}]}$ 
nearest to the origin in the complex $p$ plane are 
at $p_{[sq,1_{sd}],np} = e^{\pm \pi i/3}=(1/2)(1\pm i\sqrt{3})$ with 
unit magnitude, which is the radius of convergence of
the small-$p$ series for $\langle k \rangle_{[sq,2_{sd}]}$. 
In accord with Eqs. (\ref{ksqLyp_general})-(\ref{ksqLyp_general_denr}), 
the poles of $\langle k \rangle_{[sq,1_{sd}]}$ in the $r$ plane are the same. 


\subsection{$2_{sd}$ Square-Lattice Strip}
\label{sqd2_section}

Here, for the $2_{sd}$ strip of the square lattice, we calculate
\begin{widetext}
\beqs
\langle k \rangle_{[sq,2_{sd}]} &=& 
\frac{(1-p)^3(2-2p-4p^2+15p^3-17p^4+p^5+24p^6-34p^7+24p^8-10p^9+2p^{10})}
{2(1-2p+9p^3-18p^4+16p^5+5p^6-32p^7+44p^8-35p^9+18p^{10}-6p^{11}+p^{12})}
\cr\cr\cr
&=& \frac{r^3(1-3r^2+9r^3-2r^4-19r^5+38r^6-38r^7+24r^8-10r^9+2r^{10})}
{2(1-2r+9r^3-18r^4+16r^5+5r^6-32r^7+44r^8-35r^9+18r^{10}-6r^{11}+r^{12})} \ . 
\cr\cr
&&
\label{ksq2sd}
\eeqs
\end{widetext}
At $p=p_{c,sq}=1/2$, $\langle k \rangle_{[sq,2_{sd}]}$ has the value
\beq
\langle k \rangle_{[sq,2_{sd}]}|_{p=p_{c,sq}} = \frac{17}{118} =
0.144068 \ .
\label{ksq2sd_pc}
\eeq
The comparison of this with the value of $\langle k \rangle_{sq,c}$ for the
infinite square lattice is indicated by the ratio
\beq
R_{[sq,2_{sd}],c} = 1.4689372 \ . 
\label{ksq2sd_crit_ratio}
\eeq

For $a_{2,[sq,2_{sd}]}$ we calculate
\beq
a_{2,[sq,2_{sd}]} = \frac{235936}{107911} = 2.186394 \ . 
\label{a2_sq2sd}
\eeq

The first few terms of the small-$p$ and small-$r$ series expansions of
$\langle k \rangle_{[sq,2_{sd}]}$ are given in Table \ref{series_table}. 
In $\langle k \rangle_{[sq,2_{sd}]}$, the nearest pole to the origin in the
complex $p$ and $r$ plane is 
\beq
p_{[sq,2_{sd}],np} = r_{[sq,2_{sd}],np} = -0.4836567 \ , 
\label{ppole_sq2dual_np}
\eeq
which sets the radius of convergence of the small-$p$ and small-$r$ series
expansions for $\langle k \rangle_{[sq,2_{sd}]}$.


\subsection{$3_{sd}$ Square-Lattice Strip}
\label{sqd3_section}

For the $3_{sd}$ strip of the square lattice, we calculate
\beq
\langle k \rangle_{[sq,3_{sd}]} = \frac{N_{[sq,3_{sd}]}}{D_{[sq,3_{sd}]}} \ ,
\label{ksq3sd}
\eeq       
where $N_{[sq,3_{sd}]}$ and $D_{[sq,3_{sd}]}$ are given in Eqs.
(\ref{ksq3sd_numerator}) and (\ref{ksq3sd_denominator}) in the appendix. 
At $p=p_{c,sq}=1/2$, this has the value
\beq
\langle k \rangle_{[sq,3_{sd}]}|_{p=p_{c,sq}} = \frac{2051}{15474} =
0.132545 \ .
\label{ksq3sd_pc}
\eeq

For $a_{2,[sq,3_{sd}]}$ we calculate
\beq
a_{2,[sq,3_{sd}]} = \frac{4105669781114576}{1664338698530559} = 2.4668475  \ . 
\label{a2_sq3sd}
\eeq

The poles in 
$\langle k \rangle_{[sq,3_{sd}]}$, nearest to origin in the
complex $p$ and $r$ plane are the complex-conjugate pair
\beq
p_{[sq,3_{sd}],np} = r_{[sq,3_{sd}],np} = -0.341129 \pm 0.289364i \ , 
\label{ppole_sq3dual_np}
\eeq
with magnitude
\beq
|p_{[sq,3_{sd}],np}| = |r_{[sq,3_{sd}],np}| = 0.447326 \ , 
\label{ppole_sq3sd_np}
\eeq
which is thus the radius of convergence of the
small-$p$ and small-$r$ series expansions of $\langle k \rangle_{[sq,3_{sd}]}$.

Thus, the square-lattice strips with periodic transverse boundary
conditions and the square-lattice strips with self-dual transverse
boundary conditions most closely replicate the properties of the
infinite square lattice, namely absence of boundary effects and
self-duality.  For this reason, one expects that for a given width
$L_y$, the values of $\langle k \rangle$ and its critical value at
$p=p_{c,sq}$, $\langle k \rangle_{[sq,(L_y)_P]} |_{p=p_{sq,c}}$ or
$\langle k \rangle_{[sq,(L_y)_{sd}]} |_{p=p_{sq,c}}$ will be closer to
the values on the infinite square lattice than is the case for free
transverse boundary conditions, and our exact results confirm this
general expectation.


\section{Triangular-Lattice Strips}
\label{tri_section}


\subsection{$2_F$ Triangular-Lattice Strips}
\label{tri2f_section}

We denote the average cluster number for the strip of the triangular lattice
with width $L_y$ and free transverse boundary conditions as
$\langle k \rangle_{[tri,(L_y)_F]}$. In \cite{pc} we gave
$\langle k \rangle_{[tri,2_F]}$, which is
\beq
\langle k \rangle_{[tri,2_F]} = \frac{(1-p)^3}{1-p+p^2} =
\frac{r^3}{1-r+r^2} \ .
\label{ktri2f}
\eeq
The average cluster number is \cite{pc}
\beq
\langle k \rangle_{[tri,2_F]} = \langle k \rangle_{[sq,1_{sd}]} \ .
\label{ktri2f_eq_ksq1dual}
\eeq

In Ref. \cite{pc} we gave a numerical evaluation of this for 
$p=p_{c,tri}$, the critical bond occupation probability on the infinite
triangular lattice. Here we present an analytic expression, 
\beqs
 \langle k \rangle_{[tri,2_F]}|_{p=p_{c,tri}} &=&
\frac{2(1-6s+6s^2)}{1-2s+4s^2} \cr\cr
&=& 0.359575 \ .
\label{ktri2f_pc}
\eeqs
where here and below, we use the symbol $s=\sin(\pi/18)$, as defined
in Eq. (\ref{s}).  Owing to the equality (\ref{ktri2f_eq_ksq1dual}),
the small-$p$ and small-$r$ expansions of $\langle k
\rangle_{[tri,2_F]}$ are the same as those of $\langle k
\rangle_{[sq,1_{sd}]}$, as are the sets of poles.


\subsection{$3_F$ Triangular-Lattice Strips}
\label{tri3f_section}

For the $3_F$ strip of the triangular lattice, we calculate
\begin{widetext}
\beqs
\langle k \rangle_{[tri,3_F]}
&=& \frac{(1-p)^4(3+2p-3p^2-14p^3+48p^4-62p^5+7p^6
 +90p^7-144p^8+123p^9-66p^{10}+21p^{11}-3p^{12})}
{3(1-p-2p^3+22p^4-56p^5+72p^6-29p^7-76p^8+179p^9-210p^{10}+166p^{11}
-94p^{12}+37p^{13}-9p^{14}+p^{15})}
\cr\cr\cr
&=& \frac{r^4(2+2r+r^2-r^3-4r^4+2r^5+7r^6-27r^8+42r^9-33r^{10}+15r^{11}
-3r^{12}}
{3(1-r+r^2-2r^3+2r^4+7r^5-17r^6+22r^7-28r^8+29r^9-12r^{10}-13r^{11}+23r^{12}
  -16r^{13}+6r^{14}-r^{15})} \ .
\cr\cr
&&
\label{ktri3f}
\eeqs
\end{widetext}

At $p=p_{c,tri}$, 
\beqs
\langle k \rangle_{[tri,3_F]}|_{p=p_{c,tri}} &=&
\frac{306241-2163343s+2302182s^2}{3(25781-182124s+193812s^2)} \cr\cr
&=& 0.2714866 \ .
\label{ktri3f_pc}
\eeqs
In obtaining this and analytic evaluations of $\langle k
\rangle_{[\Lambda,(L_y)_{BC_y}]}|_{p=p_{c,\Lambda}}$ for other strips of
the triangular lattice, and for strips of the honeycomb lattice, we
have used the trigonometric identity $\sin^3(\pi/18) =
(1/8)[6\sin(\pi/18)-1]$, which enables us to reduce any
(finite-degree) polynomial in $s$ to a polynomial of degree 2.

In Table \ref{series_table} we list the first few terms in the
small-$p$ and small-$r$ series expansions of $\langle k
\rangle_{[tri,3_F]}$.  The poles in $\langle k \rangle_{[tri,3_F]}$
nearest to the origin in the complex $p$ plane are the
complex-conjugate pair
\beq
p_{[tri,3_F],np} = -0.300743 \pm 0.259341i \ ,
\label{ppole_tri3f}
\eeq
with magnitude
\beq
|p_{[tri,3_F],np}| = 0.397120 \ ,
\label{ppole_tri3f_mag}
\eeq
which sets the radius of convergence of the small-$p$ series for
$\langle k \rangle_{[tri,3_F]}$.
The pole in $\langle k \rangle_{[tri,3_F]}$ nearest to the origin in the
complex $r$ plane occurs at
\beq
r_{[tri,3_F],np} = -0.599392 \ , 
\label{rpole_tri3f}
\eeq
which sets the radius of convergence of the small-$r$ series for
$\langle k \rangle_{[tri,3_F]}$ as 0.599392. These values are listed in 
Table \ref{pole_table}. 


\subsection{$4_F$ Triangular-Lattice Strips}
\label{tri4f_section}

For the $4_F$ strip of the triangular lattice, we calculate
\beq
\langle k \rangle_{[tri,4_F]} = \frac{N_{[tri,4_F]}}{D_{[tri,4_F]}} \ ,
\label{ktri4f}
\eeq
where the numerator and denominator polynomials 
are given in Eqs. (\ref{ktri4f_numerator}) and
(\ref{ktri4f_denominator}) in the appendix. 

At $p=p_{c,tri}$,
\begin{widetext}
\beqs
\langle k \rangle_{[tri,4_F]}|_{p=p_{c,tri}} &=&
\frac{7325865108433807-51751213463154938s+55072491066145656s^2}
{8(225167815542115-1590625477629565s+1692708277627262s^2)} \cr\cr
&=& 0.229460 \ .
\label{ktri4f_pc}
\eeqs
\end{widetext}

The pole in $\langle k \rangle_{[tri,4_F]}$ nearest to the origin in the
complex $p$ plane occurs at
\beq
p_{[tri,4_F],np} = -0.335309 \ ,
\label{ppole_tri4f}
\eeq
with magnitude $|p_{[tri,4_F],np}| = 0.335309$, 
which sets the radius of convergence of the small-$p$ series for
$\langle k \rangle_{[tri,4_F]}$.
The poles in $\langle k \rangle_{[tri,4_F]}$ nearest to the origin in the
complex $r$ plane are the complex-conjugate pair 
\beq
r_{[tri,4_F],np} = -0.419061 \pm 0.379572i \ ,
\label{rpole_tri4f}
\eeq
with magnitude
\beq
|r_{[tri,4_F],np}| = 0.565408 \ , 
\label{rpole_tri4f_mag}
\eeq
which sets the radius of convergence of the small-$r$ series for
$\langle k \rangle_{[tri,4_F]}$. In contrast to the situation with the 
$2_F$ and $3_F$ strips of the triangular lattice, $|p_{[tri,4_F],np}| <
p_{tri,c}$ and $|r_{[tri,4_F],np}| < r_{tri,c}$. Thus, for this strip,
the radii of convergence of the small-$p$ and small-$r$ series are not
set by the respective physical critical values $p_{c,tri}$ and $r_{c,tri}$,
on the infinite triangular lattice, but instead by unphysical singularities. 


\subsection{$2_P$ Triangular-Lattice Strip}
\label{tri2p_section}

We denote the average cluster number for the infinite-length strip of
the triangular lattice with width $L_y$ and periodic transverse
boundary conditions as $\langle k \rangle_{[tri,(L_y)_P]}$. In
Ref. \cite{pc} we calculated $\langle k \rangle_{[tri,2_P]}$, which is
\begin{widetext}
\beqs
\langle k \rangle_{[tri,2_P]} &=& \frac{(1-p)^4(2+2p-7p^2+4p^3-p^4+2p^5-p^6)}
{2(1-2p^2+8p^3-12p^4+8p^5-2p^6)} \cr\cr\cr
&=& \frac{r^4(1+4r^2-6r^4+4r^5-r^6)}{2(1-2r^4+4r^5-2r^6)} \ . 
\label{ktri2p}
\eeqs
\end{widetext}
Here we present a simplified analytic expression for 
$\langle k \rangle_{[tri,2_P]}$ evaluated at $p=p_{c,tri}$. We find 
\beqs
\langle k \rangle_{[tri,2_P]}|_{p=p_{c,tri}} &=&
\frac{3(251-1774s+1888s^2)}{2(33-240s+256s^2)} \cr\cr
&=& 0.190910 \ .
\label{ktri2p_pc}
\eeqs

The pole in $\langle k \rangle_{[tri,2_P]}$ nearest to the origin in the
complex $p$ plane occurs at
\beq
p_{[tri,2_P],np} = -0.374357 \ , 
\label{ppole_tri2p}
\eeq
which sets the radius of convergence of the small-$p$ Taylor series for 
$\langle k \rangle_{[tri,2_P]}$ as 0.374357. 
The pole in $\langle k \rangle_{[tri,2_P]}$ nearest to the origin in the
complex $r$ plane occurs at
\beq
r_{[tri,2_P],np} = -0.6538705 \ , 
\label{rpole_tri2p}
\eeq
which sets the radius of convergence of the small-$r$ series for
$\langle k \rangle_{[tri,2_P]}$ as 0.6538705.


\subsection{$3_P$ Triangular-Lattice Strip}
\label{tri3p_section}

For the $3_P$ strip of the triangular lattice, we calculate
\beq
\langle k \rangle_{[tri,3_P]} = \frac{N_{[tri,3_P]}}{D_{[tri,3_P]}} \ , 
\label{ktri3p}
\eeq
where the numerator and denominator polynomials
are given in Eqs. (\ref{ktri3p_numerator}) and
(\ref{ktri3p_denominator}) in the appendix. 
At $p=p_{c,tri}$, 
\begin{widetext}
\beq
\langle k \rangle_{[tri,3_P]}|_{p=p_{c,tri}} =
\frac{2(74704191-527723687s+561591818s^2)}{9(939965-6640082s+7066228s^2)}
= 0.146651 \ . 
\label{ktri3p_pc}
\eeq
\end{widetext}

The poles in $\langle k \rangle_{[tri,3_P]}$ nearest to the origin in the
complex $p$ plane are the complex-conjugate pair
\beq
p_{[tri,3_P],np} = -0.2277805 \pm 0.175218i \ , 
\label{ppole_tri3p}
\eeq
with magnitude
\beq
|p_{[tri,3_P],np}| = 0.287376 \ , 
\label{ppole_tri3p_mag}
\eeq
which sets the radius of convergence of the small-$p$ series for
$\langle k \rangle_{[tri,3_P]}$.
The pole in $\langle k \rangle_{[tri,3_P]}$ nearest to the origin in the
complex $r$ plane is
\beq
r_{[tri,3_P],np} = -0.594760 \ , 
\label{rpole_tri3p}
\eeq
which sets the radius of convergence of the small-$r$ series for
$\langle k \rangle_{[tri,3_P]}$ as 0.594760.  These radii of convergence
are both smaller than the respective critical values $p_{c,tri}$ and
$r_{c,tri}$.


\subsection{$4_P$ Triangular-Lattice Strip}
\label{tri4p_section}

For the $4_P$ triangular lattice strip, we calculate
\beq
\langle k \rangle_{[tri,4_P]} = \frac{N_{[tri,4_P]}}{D_{[tri,4_P]}} \ , 
\label{ktri4p}
\eeq
where $N_{[tri,4_P]}$ and $D_{[tri,4_P]}$ are given in
Eqs. (\ref{ktri4p_numerator}) and (\ref{ktri4p_denominator}) in the
appendix.  At $p=p_{c,tri}$, 
\begin{widetext}
\beqs
\langle k \rangle_{[tri,4_P]}|_{p=p_{c,tri}} &=&
\frac{574004215646387707017-4054867821476682227104s+
  4315100205943310268010s^2}
     {2(8584252854733404261-60640688209720609514s+64532472500426786720s^2)}
\cr\cr
&=& 0.131378 \ . 
\label{ktri4p_pc}
\eeqs
\end{widetext}
Relative to the critical value, $\langle k \rangle_{c,tri}$ for the infinite
triangular lattice, 
\beq
R_{[tri,4_P,c]} = 1.174651 \ .
\label{ktri4p_ratio}
\eeq
Interestingly, this ratio is approaching reasonably close to unity
already when the strip width has the modest value of $L_y=4$, if one
uses periodic transverse boundary conditions.  The approach to the
infinite-width limit is slower if one uses free transverse boundary
conditions.  This is similar to the behavior that we found for the
square lattice, and, as in that case, one can understand it as a
consequence of the absence of any boundaries for PBC$_y$.

The pole in $\langle k \rangle_{[tri,4_P]}$ nearest to the origin in the
complex $p$ plane occurs at
\beq
p_{[tri,4_P],np} = -0.260779 \ , 
\label{ppole_tri4p}
\eeq
which sets the radius of convergence of the small-$p$ series for
$\langle k \rangle_{[tri,4_P]}$ as 0.260779. 
The pole in $\langle k \rangle_{[tri,4_P]}$ nearest to the origin in the
complex $r$ plane occurs at
\beq
r_{[tri,4_P],np} = -0.570571 \ , 
\label{rpole_tri4p}
\eeq
which sets the radius of convergence of the small-$r$ series for
$\langle k \rangle_{[tri,4_P]}$ as 0.570571. Both of these radii of convergence
are smaller than the respective critical values $p_{c,tri}$ and $r_{c,tri}$.


\section{Honeycomb-Lattice Strips}
\label{hc_section}


\subsection{$2_F$ Honeycomb-Lattice Strip}
\label{hc2f_section}

We calculated the average cluster number for the infinite-length $2_F$
strip of the honeycomb lattice in Ref. \cite{pc}
\beqs
\langle k \rangle_{[hc,2_F]} &=& \frac{(1-p)^2(4+3p+2p^2+p^3-4p^4)}
  {4(1-p^4+p^5)} \cr\cr\cr
  &=& \frac{r^2(6+6r-19r^2+15r^3-4r^4)}{4(1-r+4r^2-6r^3+4r^4-r^5)} \ .
\cr\cr
&&
\label{khc2f}
\eeqs
At $p=p_{c,hc}=1-2\sin(\pi/18) \equiv 1-2s$, this has the value
\beqs
\langle k \rangle_{[hc,2_F]}|_{p=p_{c,hc}} &=&
\frac{-55+392s-408s^2}{8(5-32s+34s^2)} \cr\cr
&=& 0.204751 \ . 
\label{khc2f_pc}
\eeqs

The first few terms in the small-$p$ and small-$r$ series expansions of 
$\langle k \rangle_{[hc,2_F]}$ are given in Table \ref{series_table}. 
The pole in $\langle k \rangle_{[hc,2_F]}$ nearest to the origin in the
complex $p$ plane occurs at
\beq
p_{[hc,2_F],np} = -0.856675 \ , 
\label{ppole_hc2f}
\eeq
which sets the radius of convergence of the small-$p$ series of
$\langle k \rangle_{[hc,2_F]}$ as 0.856675 (larger than
$p_{c,hc}=0.652704$).  The poles in $\langle k \rangle_{[hc,2_F]}$
nearest to the origin in the complex $r$ plane are the
complex-conjugate pair
\beq
r_{[hc,2_F],np} = -0.0783889 \pm 0.496940i \ , 
\label{rpole_hc2f}
\eeq
with magnitude
\beq
|r_{[tri,2_F],np}| = 0.503084 \ , 
\label{rpole_hc2f_mag}
\eeq
which sets the radius of convergence of the small-$r$ series for
$\langle k \rangle_{[hc,2_F]}$. 


\subsection{$3_F$ Honeycomb-Lattice Strip}
\label{hc3f_section}

For the $3_F$ strip of the honeycomb lattice we calculate
\begin{widetext}
\beqs
\langle k \rangle_{[hc,3_F]} &=&
\frac{(1-p)^2(3+2p-2p^2-5p^3-p^4+5p^5+p^6+4p^7-10p^8+7p^{10}-3p^{11})}
     {3(1-p^2+p^3)(1-2p^3+p^4+2p^6-2p^8+p^9)} \cr\cr\cr
     &=& \frac{r^2(1+5r-4r^2+14r^3-41r^4+87r^5-167r^6+226r^7-190r^8+95r^9
       -26r^{10}+3r^{11})}
     {3(1-r+2r^2-r^3)(1-3r+10r^2-14r^3+17r^4-26r^5+30r^6-20r^7+7r^8-r^9)} \ . 
\cr\cr
&&
\label{khc3f}
\eeqs
\end{widetext}
At $p=p_{c,hc}$, this has the value
\beqs
\langle k \rangle_{[hc,3_F]}|_{p=p_{c,hc}} &=&
\frac{-12803+90443s-96244s^2}{12(772-5453s+5803s^2)} \cr\cr
&=& 0.160002 \ . 
\label{khc3f_pc}
\eeqs

The poles in $\langle k \rangle_{[hc,3_F]}$ nearest to the origin in the
complex $p$ plane are the complex-conjugate pair
\beq
p_{[hc,3_F],np} = -0.492595 \pm 0.542272i \ , 
\label{ppole_hc3f}
\eeq
with magnitude
\beq
|p_{[hc,3_F],np}| = 0.732604 \ , 
\label{ppole_hc3f_mag}
\eeq
which sets the radius of convergence of the small-$p$ series for
$\langle k \rangle_{[hc,3_F]}$.
The poles in $\langle k \rangle_{[hc,3_F]}$ nearest to the origin in the
complex $r$ plane are the complex-conjugate pair 
\beq
r_{[hc,3_F],np} = 0.123348 \pm 0.377252i \ , 
\label{rpole_hc3f}
\eeq
with magnitude
\beq
|r_{[hc,3_F],np}| = 0.396906 \ ,
\label{rpole_hc3f_mag}
\eeq
which sets the radius of convergence of the small-$r$ series for
$\langle k \rangle_{[hc,3_F]}$. 


\subsection{$4_F$ Honeycomb-Lattice Strip}
\label{hc4f_section}

For the $4_F$ strip of the honeycomb lattice, we calculate
\beq
\langle k \rangle_{[hc,4_F]} = \frac{N_{[hc,4_F]}}{D_{[hc,4_F]}} \ , 
\label{khc4f}
\eeq
where $N_{[hc,4_F]}$ is a polynomial of degree 72 in $p$ containing a factor of
$(1-p)^2$ and $D_{[hc,4_F]}$ is a polynomial of degree 71 in $p$ that we
have calculated. At $p=p_{c,hc}$, 
\begin{widetext}
\beqs
&&\langle k \rangle_{[hc,4_F]}|_{p=p_{c,hc}}= \cr\cr
&=& \frac{-113592578275136635723243683+8024381665694504094459981670s
  -8539368606495326857081040364s^2}
     {16(53721138617890198824050135-379495673336597286155883324s
       +403850860315586504368203856s^2)}
       \cr\cr
&=& 0.1383407 \ . 
\label{khc4f_pc}
\eeqs
\end{widetext}

The poles in $\langle k \rangle_{[hc,4_F]}$ nearest to the origin in the
complex $p$ plane are the complex-conjugate pair 
\beq
p_{[hc,4_F],np} = -0.552838 \pm 0.373251i \ , 
\label{ppole_hc4f}
\eeq
with magnitude 
\beq
|p_{[hc,4_F],np}| = 0.667042 \ , 
\label{ppole_hc4f_mag}
\eeq
which sets the radius of convergence of the small-$p$ series for
$\langle k \rangle_{[hc,4_F]}$. 
The poles in $\langle k \rangle_{[hc,4_F]}$ nearest to the origin in the
complex $r$ plane are the complex-conjugate pair 
\beq
r_{[hc,4_F],np} = -0.212449 \pm 0.136692i \ , 
\label{rpole_hc4f}
\eeq
with magnitude
\beq
|r_{[tri,4_F],np}| = 0.252625 \ , 
\label{rpole_hc4f_mag}
\eeq
which sets the radius of convergence of the small-$r$ series for
$\langle k \rangle_{[hc,4_F]}$.


\subsection{$2_P$ Honeycomb-Lattice Strip}
\label{hc2p_section}

Strips of the honeycomb lattice require that $L_y$ be even.  For the
$2_P$ strip of the honeycomb lattice we calculate
\beq
\langle k \rangle_{[hc,2_P]} =  \langle k \rangle_{[sq,2_F]}  
\label{khc2p}
\eeq
For the value evaluated at $p=p_{c,hc}$, we find
\beq
\langle k \rangle_{[hc,2_P]} = \frac{-3+22s-20s^2}{4(1-2s)^2} = 0.127450 \ .
\label{khc2p_pc}
\eeq
%


\subsection{$4_P$ Honeycomb-Lattice Strip}
\label{hc4p_section}

For the $4_P$ honeycomb strip, we calculate
\beq
\langle k \rangle_{[hc,4_P]} = \frac{N_{[hc,4_P]}}{D_{[hc,4_P]}} \ ,
\label{khc4p}
\eeq
where $N_{[hc,4_P]}$ and $D_{[hc,4_P]}$ are given in Eqs.
(\ref{khc4p_numerator}) and (\ref{khc4p_denominator}) in the appendix. 
At $p=p_{c,hc}$, this has the value
\begin{widetext}
\beqs
\langle k \rangle_{[hc,4_P]}|_{p=p_{c,hc}} &=&
\frac{736538075855-5203035904036s+5536955158472s^2}
     {32(-19547696983+138088406531s-146950612867s^2)} \cr\cr
  &=& 0.0898337 \ .
\label{khc4p_pc}
\eeqs
\end{widetext}

The pole in $\langle k \rangle_{[hc,4_P]}$ nearest to the origin in the
complex $p$ plane is
\beq
p_{[hc,4_P],np} = -0.585767 \ , 
\label{ppole_hc4p}
\eeq
which sets the radius of convergence of the small-$p$ series for
$\langle k \rangle_{[hc,4_P]}$ as 0.585767. 
The pole in $\langle k \rangle_{[hc,4_P]}$ nearest to the origin in the
complex $r$ plane is 
\beq
r_{[hc,4_P],np} =-0.270891 \ ,
\label{rpole_hc4p}
\eeq
which sets the radius of convergence of the small-$r$ series for
$\langle k \rangle_{[hc,4_P]}$ as 0.270891. 


\section{Comparative Discussion}
\label{comparison_section}

As noted in the introduction, our main new results include (i) the
theorem (\ref{kform}), showing that the average cluster number per
site on infinite-length lattice strips with width $L_y$ and specified
transverse boundary conditions $BC_y$, $\langle k
\rangle_{[\Lambda,(L_y)_{BC_y}]}$, is a rational function of the bond
occupation probability $p$; (ii) the calculation of the exact
expressions for $\langle k \rangle_{[\Lambda,(L_y)_{BC_y}]}$ as a
function of $p$; (iii) exact values of these average cluster numbers
at $p=p_{c,\Lambda}$, the critical bond occupation probability for the
corresponding infinite-length lattices; (iv) a study of the $L_y$
dependence of these values (discussed further below); (v) calculations
of $d^j \langle k \rangle_{[\Lambda,(L_y)_{BC_y}]}/(dp)^j$ with
$j=1,2,3$, evaluated at $p=p_{c,\Lambda}$, for infinite-length lattice
strips $\Lambda_s$ with a resultant determination of coefficients in
the expansion of $\langle k \rangle_{[\Lambda,(L_y)_{BC_y}]}$ in
Eq. (\ref{kcrit_expansion}); and (vi) a study of the poles in $\langle
k \rangle_{[\Lambda,(L_y)_{BC_y}]}$ and the insight that these yield
concerning the role of unphysical singularities setting the radii of
convergence in small-$p$ and small-$r$ series expansions of various
quantities in percolation on infinite two-dimensional lattices.
That is, one encounters this phenomenon even for finite-width strips
of modest widths, before the limit $L_y \to \infty$ is taken to obtain
$\langle k \rangle_{\Lambda}$. 

Here we give some further comparative discussion of these results.
First, our exact results strengthen and extend two monotonocity relations
that we found in our previous study \cite{pc}. We find that for fixed
$p \in (0,1)$, $\langle k \rangle_{[\Lambda,(L_y)_{BC_y}]}$ is a
monotonically decreasing function of the strip width $L_y$ for all of
the lattices considered here.  (At the endpoints of the physical
interval in $p$, the values are fixed, as $\langle k
\rangle_{[\Lambda,(L_y)_{BC_y}]}=1$ at $p=0$ and $\langle k
\rangle_{[\Lambda,(L_y)_{BC_y}]}=0$ at $p=1$, independent of $L_y$.)
Second, for fixed $L_y$, $\langle k \rangle_{[\Lambda,(L_y)_{BC_y}]}$ is
a monotonically decreasing function of $p$ in the physical interval $0
\le p \le 1$.

Furthermore, with our present exact analytic results, we have
strengthened the finding from our previous study in \cite{pc}, that
for a given lattice type and set of transverse boundary conditions,
over the range of strip widths $L_y$ that we have studied, the
behavior of $\langle k \rangle_{[\Lambda,(L_y)_{BC_y}]}$ is consistent
with the inference that, for a fixed $p \in (0,1)$, the average
cluster number on the infinite-length strip, $\langle k
\rangle_{[\Lambda,(L_y)_{BC_y}]}$, approaches $\langle k
\rangle_{\Lambda}$ as $L_y \to \infty$.  (This is automatic for the
two endpoints, $p=0$ and $p=1$, where $\langle k \rangle_{\Lambda_s} =
1$ and $\langle k \rangle_{\Lambda_s} = 0$.)

In particular, for each type of infinite-length, finite-width lattice
strip $\Lambda_s$ for which we have calculated exact expressions for
$\langle k \rangle_{\Lambda_s}$, as $L_y$ increases, the evaluation
with $p$ set equal to the critical bond occupation probability for the
corresponding infinite two-dimensional lattice $\Lambda$,
$p=p_{c,\Lambda}$, approaches the known critical value for the
infinite lattice $\langle k \rangle_{\Lambda}|_{p=p_{c,\Lambda}}$.  As
expected, for a given width, $L_y$, the deviation from this critical
value for the infinite two-dimensional lattice is smallest for the
infinite-length strips with periodic transverse boundary conditions,
since these remove boundary effects, as contrasted to the strips with
free transverse boundary conditions:
\begin{widetext}
\beq
\Big |\langle k \rangle_{[\Lambda,(L_y)_P]}|_{p=p_{c,\Lambda}}-
 \langle k \rangle_{\Lambda}|_{p=p_{c,\Lambda}} \Big | <
\Big |\langle k \rangle_{[\Lambda,(L_y)_F]}|_{p=p_{c,\Lambda}}-
  \langle k \rangle_{\Lambda}|_{p=p_{c,\Lambda}} \Big | \ . 
\label{pfdeviation}
\eeq
\end{widetext}
Thus, with periodic transverse boundary conditions, the only
finite-size effect that remains on the infinite-length lattices is the
fact that $L_y$ is finite, i.e., there is a finite-length path
crossing the lattice strip in a transverse direction.
For a given infinite-length square-lattice strip of width $L_y$, the
deviation of the average cluster number at $p=p_{c,sq}$ from its value
on the infinite square lattice is also smaller with self-dual boundary
conditions, as contrasted with free transverse boundary conditions:
\begin{widetext}
\beq
\Big [ \langle k \rangle_{[sq,(L_y)_{sd}]}|_{p=p_{c,sq}}-
       \langle k \rangle_{sq}|_{p=p_{c,sq}} \Big ] <
\Big [ \langle k \rangle_{[sq,(L_y)_F]}|_{p=p_{c,sq}}-
 \langle k \rangle_{sq}|_{p=p_{c,sq}} \Big ] \ .
\label{sqsddeviation}
\eeq
\end{widetext} 

Our work here is complementary to the calculation in Ref.  \cite{povolotsky} of
$\langle k \rangle_{sq,diag}|_{p=p_{c,sq}}$ on infinite-length diagonal strips
of arbitrary widths (with toroidal boundary conditions) on the square lattice,
since we calculate $\langle k \rangle_{\Lambda,(L_y)_{BC_y}}$ as a function of
$p$, not just for the single value $p=p_{c,\Lambda}$, while
Ref. \cite{povolotsky} calculates the values only at $p=p_{c,sq}$. (Another
difference is that we have also calculated exact values of $\langle k
\rangle_{\Lambda,(L_y)_{BC_y}}$ for triangular and honeycomb lattices and in
\cite{pc} for the kagom\'e lattice.)  As the strip width increases, the
approach to the value $\langle k \rangle_{c,sq}$ in Eq. (\ref{kcrit_sq}) is
comparably rapid. For example, for the index $N=3$ (corresponding to a width
across the diagonal of $3\sqrt{2}=4.243$, Ref. \cite{povolotsky} obtains
$\langle k \rangle_{sq,diag}=79/672=0.117560$, which lies between our values
$\langle k \rangle_{sq,4_P}|_{p=p_{c,sq}} = 677/5572 = 0.121500$ in
Eq. (\ref{ksq4p_pc}) and $\langle k \rangle_{sq,5_P}|_{p=p_{c,sq}} =
85013/753370=0.112844$ in Eq. (\ref{ksq5p_pc}). This is in accord with
one's expectation, since the width $3\sqrt{2}$ is intermediate between the
width $L_y=4$ and $L_y=5$.

An important result of our calculations is the comparison with the
formula for the finite-size correction to $\langle k
\rangle_{c,\Lambda}$ derived in \cite{kleban_ziff,zlk}, given above in
Eq. (\ref{finite_size_correction}), both concerning the constant
$5\sqrt{3}/24$ in the $O(1/L_y^2)$ term and concerning the
universality of this finite-size correction as regards the type of
lattice, with the geometrical factors (\ref{ctri}) and (\ref{chc})
are incorporated.  For this comparison, we list in Table \ref{b_table}
the values of $\tilde b_{\Lambda,L_y}$ that we extract from our fit to
Eq. (\ref{bform}) for the infinite-length strips of the square,
triangular, and honeycomb lattices. As is evident from this table, as
$L_y$ increases, our results approach the value $\tilde b =
5\sqrt{3}/24$ in \cite{kleban_ziff} (see also \cite{zlk}) and,
furthermore, are consistent with being equal for all three of these
types of lattices, in agreement with the universality property of this
finite-size correction. Indeed, with rather modest strip widths, we
find excellent agreement with the value of $\tilde b$ in
Eq. (\ref{finite_size_correction}).  This is a valuable 
universality check using exact results for different types of lattice
strips, including square, triangular, and honeycomb lattices.

Another interesting application of our calculations of $\langle k
\rangle_{[\Lambda,(L_y)_{BC_y}]}$ on these infinite-length lattice
strips is to investigate how the small-$p$ and small-$r$ Taylor series
expansions compare with those for the corresponding infinite
two-dimensional lattices.  The entries in Table \ref{series_table} are
useful for this purpose.  As is evident from this table, for the
infinite-length $[\Lambda,(L_y)_P]$ strips, which are
$\Delta$-regular, we find that the small-$p$ expansions have the
general form $\langle k \rangle_{\Lambda_s} = 1 - (\Delta/2)p + ...$,
where the $...$ denote higher-order terms, in accord with Eq.
(\ref{dkdp_p0}). This form for the first two terms 
is the same as with the infinite two-dimension lattices.  
For the infinite-length strips that are not
$\Delta$-regular, such as those with free transverse boundary
conditions, we find that the small-$p$ expansion has the form
$\langle k \rangle_{\Lambda_s} = 1 - (\Delta_{\rm{eff}}/2)p + ...$,
where $\Delta_{\rm{eff}}$ was defined in Eq. (\ref{delta_eff}). 

Concerning the rest of the small-$p$ series, by inspecting the series
for $\langle k \rangle_{[\Lambda,(L_y)_{BC_y}]}$ on infinite-length
lattice strips of a given lattice $\Lambda$ with some specified
tranverse boundary conditions, one can see how, as a function of
increasing strip width $L_y$, coefficients of certain terms for these
strips approach the values that they have in the corresponding
small-$p$ or small-$r$ expansion of $\langle k \rangle_\Lambda$ on the
infinite two-dimension lattice $\Lambda$.  For example, consider the
$(L_y)_F$ strips of the square, triangular, and honeycomb lattices.
One sees that the coefficient of the respective linear terms in the
small-$p$ series expansions increase monotonically toward the
respective values 4, 6, and 3, in agreement with the discussion above.

The next higher-order term in the small-$p$ expansion of $\langle k
\rangle_{sq}$ for the infinite square lattice is $p^4$, and one can
see from Table \ref{series_table} how, as the width $L_y$ of the
$(L_y)_F$ square-lattice strips increases from 2 to 4, the coefficient
of the $p^4$ term in the series expansion of $\langle k
\rangle_{[sq,(L_y)_F]}$ increases toward 1, taking on the respective
values 1/2, 2/3, and 3/4. Similarly, the next term higher than linear
in the small-$p$ expansion of $\langle k \rangle_{tri}$ on the
infinite triangular lattice is $2p^3$, and the coefficients of the
$p^3$ terms in $\langle k \rangle_{[tri,(L_y)_F]}$ increase toward this
value, as 1, 4/3, and 3/2 with $L_y=2, \ 3, \ 4$,
respectively. Finally, the next term higher than linear in the
small-$p$ expansion of $\langle k \rangle_{hc}$ on the infinite
honeycomb lattice is $(1/2)p^6$, and the coefficients of the $p^6$
terms in $\langle k \rangle_{[hc,(L_y)_F]}$ increase toward 1/2, taking
on the values 1/4, 1/3, and 3/8 as $L_y$ increases from 2 to 4.

For the infinite-length strips with periodic transverse boundary
conditions, the linear terms in $p$ are equal to their values for the
corresponding infinite two-dimensional lattices, and again the rest of
the small-$p$ series become more similar to the series for the
two-dimensional lattices as the width increases.  As an example,
consider the $[sq,(L_y)_P]$ strips. The small-$p$ series for $\langle
k \rangle_{[sq,2_P]}$ has a nonzero $p^2$ term, but it is absent in the
series expansion of $\langle k \rangle_{[sq,3_P]}$ on the next wider
strip of this type. In turn, the small-$p$ series for $\langle k
\rangle_{[sq,3_P]}$ contains a nonzero $p^3$ term, but it is absent in
the series expansion of $\langle k \rangle_{[sq,4_P]}$ on the next wider
strip of this type. The small-$p$ series expansion of $\langle k
\rangle_{[sq,5_P]}$ matches not just the linear term, but also the $p^4$
term of $\langle k \rangle_{sq}$ exactly.  Corresponding comments
apply for the $(L_y)_P$ strips of the triangular and honeycomb
lattices.  One might anticipate some special properties of the
small-$p$ series expansions of $\langle k \rangle_{[sq,(L_y)_{sd}]}$
owing to the inclusion of the self-duality property. Interestingly,
one sees that with all three widths for which we have calculated
$\langle k \rangle_{[sq,(L_y)_{sd}]}$, namely, $L_y=1, \ 2, \ 3$, the
small-$p$ expansions match not just the linear term, but also the
$p^4$ term in $\langle k \rangle_{sq}$ exactly. Over this range of
$L_y$ values, one observes that the coefficient of the $p^3$ term
decreases monotonically, consistent with its vanishing as $L_y \to
\infty$.  Analogous comments apply for the small-$r$ series expansions
of $\langle k \rangle_{[\Lambda,(L_y)_{BC_y}]}$. 

Finally, we have used our exact calculations of $\langle k
\rangle_{[\Lambda,(L_y)_{BC_y}]}$ for these lattice strips to answer
an intriguing question concerning the presence of unphysical
singularities that were found, in analyses of small-$p$ and small-$r$
series calculations of average cluster numbers on two-dimensional
lattices \cite{series,domb_pearce}, to be closer to the respective
origins in these planes than the physical $p_{c,\Lambda}$ and
$r_{c,\Lambda}=1-p_{c,\Lambda}$ for these lattices.  The question is
whether such unphysical singularities (which are manifested as poles
in Pad\'e approximants of series) would also be encountered in the
exact expressions for $\langle k
\rangle_{[\Lambda,(L_y)_{BC_y}]}$. Our earlier analytic results in
\cite{pc} showed the presence of poles, but were limited to rather
narrow strip widths.  With our new calculations of $\langle k
\rangle_{\Lambda_s}$ for considerably greater strip widths, we have
answered this question, in the affirmative.  This is evident in Table
\ref{pole_table}.  Furthermore, we find that with all of the strips
for which we have performed exact calculations, for a given type of
lattice strip $\Lambda$ and specified transverse boundary conditions
$BC_y$, the magnitude of the pole(s) of $\langle k
\rangle_{[\Lambda,(L_y)_{BC_y}]}$ nearest to the origin in the complex
$p$ plane decreases monotonically with increasing $L_y$, and
similarly, the magnitude of the pole(s) of $\langle k
\rangle_{[\Lambda,(L_y)_{BC_y}]}$ nearest to the origin in the $r$
plane decreases monotonically with increasing $L_y$. Thus, the
corresponding radii of convergence of the small-$p$ and small-$r$
series also decrease with increasing $L_y$.  One knows rigorously that
the small-$p$ and small-$r$ series expansions of $\langle k
\rangle_{\Lambda}$ for infinite-length strips of arbitrarily large
width, and also for the infinite lattices $\Lambda$, are Taylor series
with finite radii of convergence, given the connection via (\ref{k})
with the Potts model. This follows because $v_p=p/(1-p)$, so that the
small-$p$ and small-$r$ expansions in this bond percolation problem
correspond, respectively, to high-temperature and low-temperature
expansions in the Potts model.  In general, the high- and
low-temperature expansions of a discrete spin model such as the Potts
model are Taylor series expansions with finite radii of convergence.
Our results are thus consistent with the inference that, as $L_y \to
\infty$, the magnitude of the pole(s) nearest to the origin in the
complex $p$ plane and the resultant radius of convergence of the
small-$p$ series expansions of $\langle k
\rangle_{[\Lambda,(L_y)_{BC_y}]}$ will approach the value obtained
from analyses of small-$p$ series expansions of $\langle k
\rangle_\Lambda$ on the corresponding infinite two-dimensional
lattices. A similar comment applies to the poles in the $r$ plane.
For example, regarding the poles in the $p$ plane, from analyses in
Ref. \cite{domb_pearce} of small-$p$ series expansions for the average
cluster number on the square lattice, $\langle k \rangle_{sq}$,
evidence was reported for an unphysical singularity at $p = -0.41 \pm
0.02$ (see also \cite{series}).  Our results, as listed in Table
\ref{pole_table} show a decrease in the magnitude of the unphysical
pole(s) nearest to the origin in the complex $p$ plane consistent with
the inference that with increasing strip width $L_y$, this magnitude
approaches this value $\simeq 0.41$ \cite{domb_pearce} obtained from
series analyses for the infinite square lattice. Indeed, the magnitude
of the complex-conjugate pair of poles nearest to the origin in the
$[sq,5_P]$ strip is already equal to 0.41 to two significant
figures. Our exact results on these lattice strips thus give a new
insight into this phenomenon of unphysical singularities closer to the
origin than $p_{c,\Lambda}$ that were noticed in early series analyses
\cite{series,domb_pearce}.


\section{Conclusions}
\label{conclusion_section} 

In this paper we have presented a number of new exact results for
average cluster numbers $\langle k \rangle_{\Lambda,(L_y)_{BC_y}}$ in
the bond percolation problem on infinite-length lattice strips of the
square, triangular, and honeycomb lattices with various transverse
boundary conditions.  We have proved a theorem that $\langle k
\rangle_{[\Lambda,(L_y)_{BC_y}]}$ is a rational function of the bond
occupation probability, $p$.  We have evaluated our expressions for
$\langle k \rangle_{[\Lambda,(L_y)_{BC_y}]}$ with $p$ set equal to the
critical values $p=p_{c,\Lambda}$ for the corresponding infinite
two-dimensional lattices.  We have also calculated coefficients of
$\langle k \rangle_{[\Lambda,(L_y)_{BC_y}]}$ in an expansion around
$p=p_{c,\Lambda}$.  Using our calculations on infinite-length strips
of several different widths and lattices types, we have checked and
found excellent agreement with the functional form and coefficient
describing the finite-size correction to the infinite-width limit.
Finally, we have carried out a study of the poles in the expressions
for $\langle k \rangle_{[\Lambda,(L_y)_{BC_y}]}$ and how these determine
the radii of convergence of the small-$p$ and small-$r$ Taylor series
expansions of these quantities. In turn, this has given a new insight 
into the appearance of unphysical singularities that were found in early 
series expansions of $\langle k \rangle_\Lambda$ on two-dimensional lattices
$\Lambda$. 


\begin{acknowledgments}

  We thank R. Ziff for informing us of Ref. \cite{povolotsky} and for
  valuable comments.  This research was supported in part by the
  Taiwan Ministry of Science and Technology grant MOST
  109-2112-M-006-008 (S.-C.C.) and by the U.S. National Science
  Foundation grant No. NSF-PHY-1915093 (R.S.).

\end{acknowledgments}


\begin{appendix}

\section{Some Detailed Results of Calculations}
\label{details_appendix}

We list here numerator and denominator polynomials in Eq. (\ref{kform})
for various infinite-length strips that are too lengthy to give in the text:
\begin{widetext}
\beqs
N_{[sq,4_F]} &=&
(1-p)^3(4+5p-13p^2-22p^3+13p^4+120p^5-35p^6-342p^7+67p^8+800p^9-42p^{10} 
-2243p^{11}
\cr\cr
&+&2042p^{12}+867p^{13}-1632p^{14}+2066p^{15}-8992p^{16}
+14900p^{17}-3933p^{18}-15767p^{19}+19105p^{20}
\cr\cr
&-&10149p^{21}+17236p^{22}-37363p^{23}+39047p^{24}
-19238p^{25}-6431p^{26}+58942p^{27}-158184p^{28}
\cr\cr
&+&235049p^{29}-176732p^{30}-19602p^{31}+213240p^{32}
-267764p^{33}+182599p^{34}-59067p^{35}-17833p^{36}
\cr\cr
&+&35509p^{37}-24007p^{38}+10257p^{39}-2997p^{40}+589p^{41}
-71p^{42}+4p^{43})
\label{ksq4f_numerator}
\eeqs
\beqs
D_{[sq,4_F]} &=& 4(1-4p^2+8p^4+21p^5-45p^6-50p^7+125p^8+106p^9-262p^{10}-388p^{11}
+1257p^{12}-911p^{13}-353p^{14}
\cr\cr
&+&1392p^{15}-3441p^{16}+7214p^{17}-7659p^{18}
-33p^{19}+10102p^{20}-13234p^{21}+12476p^{22}-17624p^{23}+25847p^{24}
\cr\cr
&-&24760p^{25}+10265p^{26}+17864p^{27}-67400p^{28}+131039p^{29}
-160372p^{30}+101976p^{31}+31616p^{32}
\cr\cr
&-&155851p^{33}+192656p^{34}-139509p^{35}+55077p^{36}+4708p^{37}
-24705p^{38}+20289p^{39}-10358p^{40}+3729p^{41}
\cr\cr
&-&961p^{42}+171p^{43}-19p^{44}+p^{45})
\label{ksq4f_denominator}
\eeqs
\beqs
N_{[sq,4_P]} &=&
(1-p)^4(4+8p-16p^3-39p^4+112p^5-20p^6-208p^7+315p^8-223p^9+248p^{10} 
-647p^{11}
\cr\cr
&+&1106p^{12}-1318p^{13}+1453p^{14}-766p^{15}-2735p^{16}
+8742p^{17}-12662p^{18}+10502p^{19}-4091p^{20}
\cr\cr
&-&1358p^{21}+3122p^{22}-2307p^{23}+1033p^{24}
-297p^{25}+51p^{26}-4p^{27})
\label{ksq4p_numerator}
\eeqs
\beqs
D_{[sq,4_P]} &=&
4(1-p+p^2)(1+p-2p^2-3p^3-3p^4+41p^5-36p^6-62p^7+140p^8-131p^9+120p^{10} 
-226p^{11}
\cr\cr
&+&460p^{12}-649p^{13}+688p^{14}-480p^{15}-654p^{16}
+3216p^{17}-5785p^{18}+5926p^{19}-3292p^{20}+99p^{21}
\cr\cr
&+&1578p^{22}-1584p^{23}+912p^{24}
-351p^{25}+90p^{26}-14p^{27}+p^{28})
\label{ksq4p_denominator}
\eeqs
\beqs
N_{[sq,5_P]} &=& (1-p)^4(5+15p+10p^2-40p^3-115p^4-29p^5+660p^6+132p^7-1709p^8
-877p^9+3950p^{10} 
\cr\cr
&+&2877p^{11}-7215p^{12}-8662p^{13}+7196p^{14}+40393p^{15}-53232p^{16}
+13204p^{17}-51313p^{18}+19634p^{19}
\cr\cr
&+&377380p^{20}-503109p^{21}-570329p^{22}+1553036p^{23}-65274p^{24}
-2873234p^{25}+4621549p^{26}
\cr\cr
&-&7720349p^{27}+15352272p^{28}-16433567p^{29}-12262362p^{30}
+78782168p^{31}-158447809p^{32}
\cr\cr
&+&214186307p^{33}-230019014p^{34}+216228871p^{35}-186980567p^{36}
+142532407p^{37}-68762291p^{38}
\cr\cr
&-&55618898p^{39}+243770621p^{40}-473742752p^{41}
+674493935p^{42}-757917965p^{43}+682330188p^{44}
\cr\cr
&-&487268491p^{45}
+263968633p^{46}-92180540p^{47}+443325p^{48}+27880668p^{49}
-24713816p^{50}+14007915p^{51}
\cr\cr
&-&5985845p^{52}+2011895p^{53}-535627p^{54}
+111561p^{55}-17625p^{56}+1993p^{57}-144p^{58}+5p^{59})
\label{ksq5p_numerator}
\eeqs
\beqs
D_{[sq,5_P]}
&=& 5(1+p-2p^2-6p^3-3p^4+26p^5+103p^6-244p^7-142p^8+516p^9+420p^{10}
-1159p^{11}-928p^{12}+1992p^{13}
\cr\cr
&+&1578p^{14}+2395p^{15}-23040p^{16}+39567p^{17}-38811p^{18}
+26672p^{19}+64051p^{20}-272943p^{21}+288026p^{22}
\cr\cr
&+&249844p^{23}-779755p^{24}+77897p^{25}+2020147p^{26}-4713372p^{27}
+8356354p^{28}-12526447p^{29}
\cr\cr
&+&9141812p^{30}+17671571p^{31}-77282022p^{32}
+157595007p^{33}-229642624p^{34}+269077829p^{35}
\cr\cr
&-&270785628p^{36}+242608396p^{37}-188685316p^{38}+99894346p^{39}
+41909295p^{40}-247536567p^{41}
\cr\cr
&+&498168300p^{42}-733592566p^{43}+871224361p^{44}
-853561993p^{45}+688825282p^{46}-448046653p^{47}
\cr\cr
&+&220169597p^{48}-63265190p^{49}-13220314p^{50}+33003789p^{51}-26681313p^{52}
+15080448p^{53}-6672527p^{54}
\cr\cr
&+&2389811p^{55}-697964p^{56}+165068p^{57}-31003p^{58}+4465p^{59}-464p^{60}
+31p^{61}-p^{62})
\label{ksq5p_denominator}
\eeqs
\beqs
N_{[sq,3_{sd}]} &=& (1-p)^3(3-3p-21p^2+37p^3+97p^4-265p^5-275p^6+1559p^7-735p^8
-4454p^9+6397p^{10}+7719p^{11}
\cr\cr
&-&25594p^{12}+461p^{13}+76993p^{14}-100105p^{15}
-48081p^{16}+240589p^{17}-133404p^{18}-299125p^{19}
\cr\cr
&+&397672p^{20}+468568p^{21}
-1660402p^{22}+1467662p^{23}+705502p^{24}-2859795p^{25}+2447284p^{26}
\cr\cr
&-&148761p^{27}+71758p^{28}-4717102p^{29}+10333853p^{30}
-9242363p^{31}-2195761p^{32}+18554630p^{33}
\cr\cr
&-&29140317p^{34}+26914438p^{35}-13774889p^{36}-2046623p^{37}
+12789267p^{38}-15764460p^{39}
\cr\cr
&+&13053019p^{40}-8354879p^{41}+4320752p^{42}-1833211p^{43}+638949p^{44}
-181329p^{45}+41088p^{46}
\cr\cr
&-&7182p^{47}+912p^{48}-75p^{49}+3p^{50})
\label{ksq3sd_numerator}
\eeqs
\beqs
D_{[sq,3_{sd}]} &=& 3(1-2p-5p^2+19p^3+13p^4-112p^5+32p^6+542p^7-883p^8
-788p^9+3568p^{10}-1056p^{11}-9489p^{12}
\cr\cr
&+&11669p^{13}+18234p^{14}-61546p^{15}+42562p^{16}+71008p^{17}
-151651p^{18}+24638p^{19}+201958p^{20}-77630p^{21}
\cr\cr
&-&616216p^{22}+1248416p^{23}-776173p^{24}-859257p^{25}
+2199256p^{26}-1891602p^{27}+827535p^{28}
\cr\cr
&-&1633704p^{29}+5205016p^{30}-8217812p^{31}+5395624p^{32}+4995137p^{33}
-18179344p^{34}+25979096p^{35}
\cr\cr
&-&23240423p^{36}+11597999p^{37}+2383494p^{38}-12270154p^{39}+15475815p^{40}
-13388744p^{41}+9080298p^{42}
\cr\cr
&-&5043522p^{43}+2332946p^{44}-902428p^{45}+290686p^{46}-77040p^{47}
+16440p^{48} -2725p^{49}+330p^{50}\cr\cr
&-&26p^{51}+p^{52})
\label{ksq3sd_denominator}
\eeqs
\beqs
N_{[tri,4_F]} &=& (1-p)^4(2-5p-4p^2+41p^3-52p^4-80p^5+164p^6+838p^7-4165p^8
+8517p^9-8197p^{10}-1589p^{11}
\cr\cr
&+&13355p^{12}-8786p^{13}-4606p^{14}-61665p^{15}+374163p^{16}
-1043384p^{17}+1905928p^{18}-2421614p^{19}
\cr\cr
&+&1878238p^{20}+140422p^{21}-3349440p^{22}+6775564p^{23}-9239709p^{24}
+9983576p^{25}-9009248p^{26}
\cr\cr
&+&6948032p^{27}-4628988p^{28}+2674993p^{29}-1339759p^{30}+578446p^{31}
-213006p^{32}+65734p^{33}-16546p^{34}
\cr\cr
&+&3259p^{35}-470p^{36}+44p^{37}-2p^{38})
\label{ktri4f_numerator}
\eeqs
\beqs
D_{[tri,4_F]} &=& 2(1-4p+4p^2+18p^3-60p^4+43p^5+80p^6+225p^7-2534p^8+8252p^9
-15122p^{10}+15527p^{11}-3457p^{12}
\cr\cr
&-&12950p^{13}+17747p^{14}-40206p^{15}+232540p^{16}-868823p^{17}
+2151018p^{18}-3857167p^{19}+5099429p^{20}
\cr\cr
&-&4529584p^{21}+1058496p^{22}+5305640p^{23}-13128301p^{24}+20075271p^{25}
-24025016p^{26}+24060724p^{27}
\cr\cr
&-&20745034p^{28}+15614712p^{29}-10330842p^{30}+6022412p^{31}
-3090155p^{32}+1389384p^{33}-542808p^{34}
\cr\cr
&+&181832p^{35}-51206p^{36}+11780p^{37}-2122p^{38}+280p^{39}-24p^{40}+p^{41})
\label{ktri4f_denominator}
\eeqs
\beqs
N_{[tri,3_P]} &=& (1-p)^6(3+9p-50p^3+84p^4-24p^5-192p^6+554p^7-844p^8+812p^9
-516p^{10}
\cr\cr
&+&246p^{11}-151p^{12}+143p^{13}-112p^{14}+56p^{15}-16p^{16}+2p^{17})
\label{ktri3p_numerator}
\eeqs
\beqs
D_{[tri,3_P]} &=& 3(1-3p^2-3p^3+68p^4-187p^5+190p^6+162p^7-1035p^8+2404p^9
-3822p^{10} \cr\cr
&+&4494p^{11}-3954p^{12}+2580p^{13}-1215p^{14}+391p^{15}-77p^{16}+7p^{17})
\label{ktri3p_denominator}
\eeqs
\beqs
N_{[tri,4_P]} &=& (1-p)^6(4+12p-4p^2-100p^3-83p^4+1290p^5-2067p^6-2512p^7
+11219p^8+3776p^9-91473p^{10}
\cr\cr
&+&237866p^{11}-238434p^{12}-355578p^{13}+2194759p^{14}-5879228p^{15}
+10734693p^{16}-11817298p^{17}-3000450p^{18}
\cr\cr
&+&49716006p^{19}-133234513p^{20}+226293288p^{21}-260526672p^{22}
+145333622p^{23}+188864004p^{24}-743143968p^{25}
\cr\cr
&+&1422696984p^{26}-2051609680p^{27}+2439158465p^{28}-2472507822p^{29}
+2176639966p^{30}-1694462238p^{31}
\cr\cr
&+&1200557665p^{32}-813001894p^{33}+559873482p^{34}-405661962p^{35}
+300606573p^{36}-214020448p^{37}
\cr\cr
&+&138935000p^{38}-79728612p^{39}+39753500p^{40}-17021640p^{41}+6188754p^{42}
-1884492p^{43}+471692p^{44}
\cr\cr
&-&94508p^{45}+14570p^{46}-1622p^{47}+116p^{48}-4p^{49})
\label{ktri4p_numerator}
\eeqs
\beqs
D_{[tri,4_P]} &=& 4(1-4p^2-8p^3+42p^4+258p^5-1514p^6+2760p^7+81p^8-7196p^9
-8065p^{10}+116560p^{11}-367969p^{12}
\cr\cr
&+&562624p^{13}+89861p^{14}-3170072p^{15}
+10610941p^{16}-22666338p^{17}+32951037p^{18}-21009590p^{19}-50850559p^{20}
\cr\cr
&+&222682606p^{21}-494752835p^{22}+772870308p^{23}
-843281180p^{24}+417584024p^{25}+751628022p^{26}
\cr\cr
&-&2729780780p^{27}+5298392040p^{28}-7950694944p^{29}
+10021670376p^{30}-10934130274p^{31}+10454310676p^{32}
\cr\cr
&-&8802418934p^{33}+6535447502p^{34}-4275633432p^{35}+2459294308p^{36}
-1239099924p^{37}+543948012p^{38}
\cr\cr
&-&206498264p^{39}+67102916p^{40}-18406832p^{41}+4181204p^{42}-765524p^{43}
+108540p^{44}
\cr\cr
&-&11180p^{45}+744p^{46}-24p^{47})
\label{ktri4p_denominator}
\eeqs
\beqs
N_{[hc,4_P]} &=& (1-p)^3(4+6p+2p^2-6p^3-20p^4+12p^5+12p^6-2p^7-11p^8
+23p^9-9p^{10}
\cr\cr
&-&46p^{11}+118p^{12}-207p^{13}+257p^{14}-159p^{15}-63p^{16}
+194p^{17}-126p^{18}+24p^{19}
\cr\cr
&-&71p^{20}+309p^{21}-623p^{22}+705p^{23}-391p^{24}-20p^{25}+178p^{26}
-116p^{27}+34p^{28}-4p^{29})
\label{khc4p_numerator}
\eeqs
\beqs
D_{[hc,4_P]} &=& 4(1-p^2-p^3-2p^4+10p^5-5p^6-3p^7+10p^9-13p^{10}-6p^{11}
+47p^{12}-105p^{13}+167p^{14}
\cr\cr
&-&182p^{15}+99p^{16}+39p^{17}-118p^{18}+95p^{19}-55p^{20}+110p^{21}
-286p^{22}+481p^{23}-515p^{24}
\cr\cr
&+&317p^{25}-43p^{26}-104p^{27}+98p^{28}-43p^{29}+10p^{30}-p^{31})
\label{khc4p_denominator}
\eeqs
\end{widetext}


\end{appendix}


\begin{table}
  \caption{\footnotesize{Structural features of exact expressions for
      average cluster numbers
      $\langle k \rangle_{[\Lambda,(L_y)_{BC_y}]}$ on infinite-length strips of
      various lattices $\Lambda$ with width $L_y$ and
      specified transverse boundary conditions,
      $(BC)_y$, expressed as functions of bond occupation probability $p$ and
      bond vacancy probability $r=1-p$. For each such lattice strip we list
      the degrees ${\rm deg}(N_{[\Lambda,(L_y)_{BC_y}}])$ and
      ${\rm deg}(D_{[\Lambda,(L_y)_{BC_y}]})$
      of the numerator and denominator of 
$\langle k \rangle_{[\Lambda,(L_y)_{BC_y}]}$ (as polynomials in $p$ or $r$), 
   and the degree
      ${\rm deg}[PF(N_{[\Lambda,(L_y)_{BC_y}]})]$ 
      To save space, in the table we write
      ${\rm deg}(N_{[\Lambda,(L_y)_{BC_y}]}) \equiv {\rm deg}(N)$,
      ${\rm deg}(D_{[\Lambda,(L_y)_{BC_y}]}) \equiv {\rm deg}(D)$, and
      ${\rm deg}[PF(N_{[\Lambda,(L_y)_{BC_y}]})] \equiv {\rm deg}[PF(N)]$.}}
\begin{center}
\begin{tabular}{|c|c|c|c|c|} \hline\hline
  $\Lambda$ & $(L_y)_{BC_y}$ & ${\rm deg}(N)$ &
  ${\rm deg}(D)$ & ${\rm deg}[PF(N)]$ \\
  \hline\hline
sq & $1_F$     & 1     & 0   &  1    \\
sq & $2_F$     & 4     & 3   &  2    \\
sq & $3_F$     & 13    & 12  &  3    \\ 
sq & $4_F$     & 46    & 45  &  3    \\
\hline
sq & $2_P$     & 5     & 4   &  2    \\
sq & $3_P$     & 11    & 10  &  3    \\
sq & $4_P$     & 31    & 30  &  4    \\
sq & $5_P$     & 63    & 62  &  4    \\
\hline
sq & $1_{sd}$  & 3     & 2   &  3    \\
sq & $2_{sd}$  & 13    & 12  &  3    \\
sq & $3_{sd}$  & 53    & 52  &  3    \\
\hline
tri& $2_F$     & 3     & 2   &  3    \\
tri& $3_F$     & 16    & 15  &  4    \\
tri& $4_F$     & 42    & 41  &  4    \\
\hline    
tri& $2_P$     & 10    & 6   &  4    \\
tri& $3_P$     & 23    & 17  &  6    \\
tri& $4_P$     & 55    & 47  &  6    \\
\hline
hc & $2_F$     & 6     & 5   &  2    \\
hc & $3_F$     & 13    & 12  &  2    \\
hc & $4_F$     & 72    & 71   & 2    \\
\hline    
hc & $2_P$     & 4     & 3   &  2    \\
hc & $4_P$     & 32    & 31  &  3    \\
\hline\hline
\end{tabular}
\end{center}
\label{kform_table}
\end{table}
%


\begin{table}
  \caption{\footnotesize{Values of average cluster numbers
      $\langle k \rangle_{[\Lambda,(L_y)_{BC_y}]}$ on infinite-length strips of
      various lattices with specified transverse boundary conditions,
      evaluated at the critical bond occupation probabilities
      $p=p_{c,\Lambda}$ for the corresponding infinite two-dimensional
      lattices. These values are given
      analytically and numerically, to the indicated floating-point accuracy.
      The entries in the right-most column of the table are the values of
      the ratio $R_{[\Lambda,(L_y)_{BC_y}]}$ in Eq. (\ref{kcrit_ratio}).}}
\begin{center}
\begin{tabular}{|c|c|c|c|c|} \hline\hline
  $\Lambda$ & $(L_y)_{BC_y}$ &
  $\langle k \rangle_{[\Lambda,(L_y)_{BC_y}]}$ &
  $\langle k \rangle_{[\Lambda,(L_y)_{BC_y}],num.}$ &
  $R_{[\Lambda,(L_y)_{BC_y}]}$ 
  \\ \hline
sq & $1_F$    & 1/2                & 0.5       & 5.098076   \\
sq & $2_F$    & 2/7                & 0.285714  & 2.913186   \\
sq & $3_F$    & 147/670            & 0.219403  & 2.237066   \\ 
sq & $4_F$    & 27229/145196       & 0.187533  & 1.912112   \\
\hline
sq & $2_P$    & 1/5                & 0.2       & 2.039230   \\
sq & $3_P$    & 11/78              & 0.141026  & 1.437919   \\
sq & $4_P$    & 677/5572           & 0.121500  & 1.238836   \\
sq & $5_P$    & 85013/753370       & 0.112844  & 1.150571   \\
\hline
sq & $1_{sd}$ & 1/6                & 0.166667  & 1.699359   \\
sq & $2_{sd}$ & 17/118             & 0.144068  & 1.468937   \\
sq & $3_{sd}$ & 2051/15474         & 0.132545  & 1.351448   \\
\hline
sq & $\infty$ &Eq.(\ref{kcrit_sq}) & 0.0980762 & 1          \\
\hline
tri & $2_F$   & Eq.  (\ref{ktri2f_pc}) & 0.359575  & 3.214963   \\
tri & $3_F$   & Eq. (\ref{ktri3f_pc})  & 0.271487  & 2.427362   \\
tri & $4_F$   & Eq. (\ref{ktri4f_pc})  & 0.229460  & 2.051605   \\
\hline    
tri & $2_P$   & Eq. (\ref{ktri2p_pc})  & 0.190910  & 1.706929   \\
tri & $3_P$   & Eq. (\ref{ktri3p_pc})  & 0.146651  & 1.311205   \\
tri & $4_P$   & Eq. (\ref{ktri4p_pc})  & 0.131378  & 1.174651   \\
\hline
tri & $\infty$ & Eq. (\ref{kcrit_tri}) & 0.111844  & 1 \\
\hline
hc & $2_F$     & Eq. (\ref{khc2f_pc})  & 0.204751  & 2.663717   \\
hc & $3_F$     & Eq. (\ref{khc3f_pc})  & 0.160002  & 2.081555   \\
hc & $4_F$     & Eq. (\ref{khc4f_pc})  & 0.138341  & 1.799749   \\
\hline
hc & $2_P$     & Eq. (\ref{khc2p_pc})  & 0.127450  & 1.658066   \\
hc & $4_P$     & Eq. (\ref{khc4p_pc})  & 0.0898337 & 1.168696   \\
\hline
hc &$\infty$   & Eq. (\ref{kcrit_hc})  & 0.076867  & 1 \\
\hline\hline
\end{tabular}
\end{center}
\label{kcrit_table}
\end{table}
%


\begin{table}
  \caption{\footnotesize{Values of $a_{i, [sq,(L_y)_{BC_y}]}$ for 
      $i=1,2$ in Eq.
      (\ref{kstrip_expansion}) for infinite-length, finite-width
      square-lattice strips with various transverse boundary conditions.}}
\begin{center}
\begin{tabular}{|c|c|c|} \hline\hline
  $(L_y)_{BC_y}$ & $a_{1,[sq,(L_y)_{BC_y}]}$ &
                   $a_{2,[sq,(L_y)_{BC_y}]}$ \\ \hline
$1_F$  & $-1$         & 0          \\
$2_F$  & $-1.204082$  & 0.921283   \\
$3_F$  & $-1.214450$  & 1.464119   \\ 
$4_F$  & $-1.200912$  & 1.833688   \\
\hline
$2_P$  & $-1$         & 1.493333   \\
$3_P$  & $-1$         & 2.201755   \\
$4_P$  & $-1$         & 2.617979   \\
$5_P$  & $-1$         & 2.898863   \\
\hline
$1_{sd}$ & $-1$       & 1.777778   \\
$2_{sd}$ & $-1$       & 2.186394   \\
$3_{sd}$ & $-1$       & 2.4668475  \\
\hline\hline
\end{tabular}
\end{center}
\label{aj_table}
\end{table}
%


\begin{table}
  \caption{\footnotesize{Small-$p$ and small-$r$ expansions of the average
    cluster number $\langle k \rangle_{[\Lambda,(L_y)_{BC_y}]}$ for the 
    infinite-length strip of the lattice  $\Lambda$ with width $L_y$ and 
    transverse boundary conditions $BC_y$.}}
\begin{center}
\begin{tabular}{|c|c|c|c|} \hline\hline
  $\Lambda$ & $(L_y)_{BC_y}$ & small-$p$ series & small-$r$ series \\ 
\hline
sq & $1_F$ & $1-p$ \ (exact)   & $r$ \ (exact)          \\
sq & $2_F$ & $1-\frac{3}{2}p+\frac{1}{2}p^4+\frac{1}{2}p^6+O(p^7)$ &
$\frac{1}{2}r^2+2r^3-\frac{7}{2}r^5-\frac{3}{2}r^6+O(r^7)$ \\
sq & $3_F$ & $1-\frac{5}{3}p+\frac{2}{3}p^4+p^6+O(p^7)$ &
$r^3+\frac{7}{3}r^4+2r^5-\frac{11}{3}r^6+O(r^7)$ \\ 
sq & $4_F$ & $1-\frac{7}{4}p+\frac{3}{4}p^4+\frac{5}{4}p^6+O(p^7)$ &
$\frac{1}{2}r^3+\frac{5}{4}r^4+2r^5+\frac{19}{4}r^6+O(r^7)$ \\ 
\hline
sq & $2_P$ & $1-2p+\frac{1}{2}p^2+2p^4-2p^5+\frac{5}{2}p^6+O(p^7)$ &
$\frac{1}{2}r^2+2r^4-2r^5+\frac{5}{2}r^6-6r^7+O(r^8)$ \\
sq & $3_P$ & $1-2p+\frac{1}{3}p^3+p^4+2p^5+O(p^7)$ &
$\frac{1}{3}r^3+r^4+2r^5-2r^7+O(r^8)$ \\ 
sq & $4_P$ & $1-2p+\frac{5}{4}p^4+5p^6+O(p^7)$     &
$\frac{5}{4}r^4+5r^6-4r^7+O(r^8)$ \\
sq & $5_P$ & $1-2p+p^4+\frac{1}{5}p^5+2p^6+O(p^7)$ &
$r^4+\frac{1}{5}r^5+2r^6+2r^7+O(r^8)$ \\
\hline
sq & $1_{sd}$ & $1-2p+p^3+p^4-p^6-p^7+O(p^9)$      &
$r^3+r^4-r^6-r^7+O(r^9)$   \\
sq & $2_{sd}$ & $1-2p+\frac{1}{2}p^3+p^4+\frac{1}{2}p^5+p^6+O(p^7)$ &
$\frac{1}{2}r^3+r^4+\frac{1}{2}r^5+r^6+r^7+O(r^8)$
  \\
sq &$3_{sd}$ & $1-2p+\frac{1}{3}p^3+p^4+\frac{1}{3}p^5+\frac{4}{3}p^6+O(p^7)$ &
$\frac{1}{3}r^3+r^4+\frac{1}{3}r^5+\frac{4}{3}r^6+\frac{1}{3}r^7+O(r^8)$ \\
\hline
sq & $\infty$& $1-2p+p^4+2p^6+O(p^7)$ &
$r^4+2r^6-2r^7+O(r^8)$ \\ 
\hline\hline
tri & $2_F$   & $1-2p+p^3+p^4-p^6+O(p^7)$    &
$r^3+r^4-r^6-r^7+r^9+r^{10}+O(r^{12})$      \\
tri & $3_F$   & $1-\frac{7}{3}p+\frac{4}{3}p^3+\frac{5}{3}p^4+p^5
-\frac{1}{3}p^6+O(p^7)$ &
$\frac{2}{3}r^4+\frac{4}{3}r^5+r^6+\frac{2}{3}r^7-\frac{1}{3}r^8
-\frac{17}{3}r^9-\frac{5}{3}r^{10}+O(r^{11})$   \\
tri & $4_F$   & $1-\frac{5}{2}p+\frac{3}{2}p^3+2p^4+\frac{3}{2}p^5
+\frac{1}{2}p^6+O(p^7)$ &
$\frac{1}{2}r^4+r^6+\frac{3}{2}r^7+2r^8+2r^9-\frac{1}{2}r^{10}+O(r^{11})$ \\ 
\hline    
tri & $2_P$   & $1-3p+\frac{1}{2}p^2+4p^3+\frac{9}{2}p^4-10p^5-10p^6+O(p^7)$  &
$\frac{1}{2}r^4+2r^6-2r^8+O(r^{10})$ \\
tri & $3_P$   & $1-3p+\frac{7}{3}p^3+6p^4+11p^5-\frac{17}{3}p^6+O(p^7)$  &
$\frac{4}{3}r^6+2r^8+2r^{10}+O(r^{11})$
\\
tri & $4_P$   & $1-3p+2p^3+\frac{13}{4}p^4+7p^5+22p^6+O(p^7)$          &
$r^6+\frac{1}{4}r^8+6r^{10}+O(r^{11})$ \\
\hline
tri &$\infty$ & $1-3p+2p^3+3p^4+3p^5+3p^6+O(p^7)$  &
$r^6+3r^{10}+3r^{11}+O(r^{12})$ \\
\hline\hline
hc & $2_F$   & $1-\frac{5}{4}p+\frac{1}{4}p^6+\frac{1}{4}p^{10}+O(p^{11})$ &
$\frac{3}{2}r^2+3r^3-\frac{31}{4}r^4-7r^5+35r^6+O(r^7)$    \\
hc & $3_F$   & $1-\frac{4}{3}p+\frac{1}{3}p^6+\frac{2}{3}p^{10}+O(p^{11})$  &
$\frac{1}{3}r^2+3r^3+\frac{17}{3}r^4-\frac{22}{3}r^5-53r^6+O(r^7)$       \\
hc & $4_F$   & $1-\frac{11}{8}p+\frac{3}{8}p^6+\frac{7}{8}p^{10}+O(p^{11})$ &
$\frac{1}{4}r^2+\frac{3}{2}r^3+\frac{29}{8}r^4+\frac{93}{8}r^5+
\frac{35}{8}r^6+O(r^7)$       \\
\hline
hc & $2_P$  & $1-\frac{3}{2}p+\frac{1}{2}p^4+\frac{1}{2}p^6-\frac{1}{2}p^7
+\frac{1}{2}p^8-p^9+p^{10}+O(p^{11})$ &
$\frac{1}{2}r^2+2r^3-\frac{7}{5}r^5-\frac{3}{2}r^6+O(r^7)$  \\
hc & $4_P$  & $1-\frac{3}{2}p+\frac{1}{2}p^6+\frac{3}{4}p^8+\frac{5}{2}p^{10}
+O(p^{11})$ &
$r^3+\frac{9}{4}r^4+\frac{11}{2}r^5+7r^6+O(r^7)$   \\ 
\hline
hc & $\infty$& $1-\frac{3}{2}p+\frac{1}{2}p^6+\frac{3}{2}p^{10}+O(p^{11})$ 
& $r^3+\frac{3}{2}r^4+\frac{3}{5}r^6+O(r^7)$ \\
\hline\hline
\end{tabular}
\end{center}
\label{series_table}
\end{table}


\begin{table}
  \caption{\footnotesize{For each infinite-length strip of the lattice
      $\Lambda$ with width $L_y$ and transverse boundary conditions
      $BC_y$, denoted $[\Lambda,(L_y)_{BC_y}]$, this table lists
      information about the pole or complex-conjugate pair of poles
      located nearest to the origin in the complex $p$ or $r$ plane,
      in the exact expression for the average cluster number, $\langle
      k \rangle_{[\Lambda,(L_y)_{BC_y}]}$. The columns are: (i)
      $\Lambda$, (ii) $(L_y)_{BC_y}$, (iii)
      $p_{[\Lambda,(L_y)_{BC_y}],np}$, (iv)
      $|p_{[\Lambda,(L_y)_{BC_y}],np}|$, (v) whether
      $|p_{[\Lambda,(L_y)_{BC_y}],np}|$ is larger or smaller than the
      critical bond occupation probability $p_{c,\Lambda}$ on the
      corresponding infinite two-dimensional lattice, (vi)
      $r_{[\Lambda,(L_y)_{BC_y}],np}$, (vii)
      $|r_{[\Lambda,(L_y)_{BC_y}],np}|$, (viii) whether
      $|r_{[\Lambda,(L_y)_{BC_y}],np}|$ is larger or smaller than the
      critical bond occupation probability $r_{c,\Lambda}$ on the
      corresponding infinite two-dimensional lattice,.  The values of
      $p_{c,\Lambda}$ are given in Eqs. (\ref{pc_sq})-(\ref{pc_hc}),
      and $r_{c,\Lambda}=1-p_{c,\Lambda}$.  For brevity of notation,
      column (v) is labelled with the symbol $rpc$, standing for
      $|p_{[\Lambda,(L_y)_{BC_y}],np}|$ ``relative to $p_{c,\Lambda}$''
      and similarly, column (viii) is labelled with the symbol $rrc$,
      standing for $|[r_{[\Lambda,(L_y)_{BC_y}],np}|$ ``relative to
      $r_{c,\Lambda}$''.  Where an entry is not applicable, we
      indicate this with $-$.}}
\begin{center}
\begin{tabular}{|c|c|c|c|c|c|c|c|} \hline\hline
  $\Lambda$ & $(L_y)_{BC_y}$ & $p_{[\Lambda,(L_y)_{BC_y}],np}$ & 
$|p_{[\Lambda,(L_y)_{BC_y}],np}|$ & $rpc$ & 
$r_{[\Lambda,(L_y)_{BC_y}],np}$ & 
$|r_{[\Lambda,(L_y)_{BC_y}],np}|$ & $rrc$  \\ \hline
  sq & $1_F$ & none                     & $-$      & $-$        &
               none                     & $-$      & $-$        \\
sq & $2_F$ & $-0.754878$              & 0.754878 & $> p_{c,sq}$ & 
             $0.122561 \pm 0.744862i$ & 0.754878 & $> r_{c,sq}$ \\
sq & $3_F$ & $-0.400758\pm 0.399068i$ & 0.565564 & $> p_{c,sq}$ & 
             $-0.411578$              & 0.411578 & $< r_{c,sq}$ \\ 
sq & $4_F$ & $-0.492588$              & 0.492588 & $< p_{c,sq}$ &
             $-0.317578\pm 0.244625i$ & 0.400871 & $< r_{c,sq}$ \\
\hline
sq & $2_P$ & $-0.618034$ & 0.618034              & $> p_{c,sq}$ &
             $-0.618034$ & 0.618034              & $> r_{c,sq}$ \\
sq & $3_P$ & $-0.354731 \pm 0.319907i$ &0.477676 & $< p_{c,sq}$ & 
             $-0.354731 \pm 0.319907i$ &0.477676 & $< r_{c,sq}$ \\
sq & $4_P$ & $-0.424294$               & 0.424294& $< p_{c,sq}$ &
             $-0.424294$               & 0.424294& $< r_{c,sq}$ \\
sq & $5_P$ & $-0.371844 \pm 0.169863i$ & 0.408805& $< p_{c,sq}$ &
             $-0.371844 \pm 0.169863i$ & 0.408805& $< r_{c,sq}$ \\
\hline
sq &$1_{sd}$ & $e^{\pm i\pi/3}$        & 1       & $> p_{c,sq}$ & 
               $e^{\pm i\pi/3}$        & 1       & $> r_{c,sq}$ \\
sq &$2_{sd}$ & $-0.483657$             & 0.483657& $< p_{c,sq}$ &
               $-0.483657$             & 0.483657& $< r_{c,sq}$ \\
sq &$3_{sd}$ & $-0.341129 \pm 0.289364i$ &0.447326&$< p_{c,sq}$ &
               $-0.341129 \pm 0.289364i$ &0.447326&$< p_{c,sq}$ \\
\hline\hline
tri& $2_F$ & $e^{\pm i\pi/3}$          & 1       & $> p_{c,tri}$ & 
             $e^{\pm i\pi/3}$          & 1       & $> r_{c,tri}$ \\
tri& $3_F$ & $-0.300743\pm 0.259341i$  & 0.397120 & $> p_{c,tri}$ &
             $-0.599392$               & 0.599392 & $< r_{c,tri}$ \\
tri& $4_F$ & $-0.335309$               & 0.335309 & $< p_{c,tri}$ &
             $-0.419061 \pm 0.379572i$ & 0.565408 & $< r_{c,tri}$ \\ 
\hline    
tri& $2_P$ & $-0.374357$               & 0.374357  & $> p_{c,tri}$ & 
             $-0.6538705$              & 0.6538705 & $> r_{c,tri}$ \\
tri& $3_P$&  $-0.2277805 \pm 0.175218i$& 0.287376  & $< p_{c,tri}$ &
             $-0.594760$               & 0.594760  & $< r_{c,tri}$ \\
tri& $4_P$ & $-0.260779$               & 0.260779  & $< p_{c,tri}$ &
             $-0.570571$               & 0.570571  & $< r_{c,tri}$ \\
\hline\hline
hc & $2_F$ & $-0.856675$               & 0.856675  & $> p_{c,hc}$ & 
             $-0.0783889\pm 0.496940i$ & 0.503084  & $> r_{c,hc}$ \\
hc & $3_F$ & $-0.492595\pm 0.542272i$  & 0.732604  & $> p_{c,hc}$ &
             $0.123348 \pm 0.377252i$  & 0.396906  & $> r_{c,hc}$ \\
hc & $4_F$ & $-0.552838 \pm 0.373251i$ & 0.667042  & $> p_{c,hc}$ &
             $-0.212449 \pm 0.136692i$ & 0.252625  & $< r_{c,hc}$ \\
\hline
hc & $2_P$ & $-0.754878$              & 0.754878   & $> p_{c,hc}$ & 
             $0.122561 \pm 0.744862i$ & 0.754878   & $> r_{c,hc}$  \\
hc & $4_P$ & $-0.585767$               & 0.585767  & $< p_{c,hc}$ &
             $-0.270891$               & 0.270891  & $< r_{c,hc}$ \\
\hline\hline
\end{tabular}
\end{center}
\label{pole_table}
\end{table}


\begin{table}
  \caption{\footnotesize{Values of $\tilde b_{[\Lambda,(L_y)_P]}$ in Eq.
      (\ref{btilde}) for infinite-length, finite-width
      lattice strips with periodic transverse boundary conditions,
      including a comparison with the value $\tilde b=5\sqrt{3}/24=
      0.360844$
      in Eq. (\ref{finite_size_correction}) from Ref.
      \cite{kleban_ziff} (see also \cite{zlk}).}}
\begin{center}
  \begin{tabular}{|c|c|c|c|} \hline\hline
    $\Lambda$ & $(L_y)_P$ & $\tilde b_{[\Lambda,(L_y)_P]}$ &
    $\frac{\tilde b_{[\Lambda,(L_y)_P]}}{\tilde b_\Lambda}$ \\ \hline
sq  & $2_P$  & 0.407695  & 1.129838    \\
sq  & $3_P$  & 0.386545  & 1.071225    \\
sq  & $4_P$  & 0.374786  & 1.038638    \\
sq  & $5_P$  & 0.369185  & 1.023116    \\
\hline
tri & $2_P$  & 0.365190  & 1.012044    \\
tri & $3_P$  & 0.361720  & 1.002428    \\
tri & $4_P$  & 0.360890  & 1.0001279   \\
\hline
hc  & $2_P$  & 0.350452  & 0.971201    \\
hc  & $4_P$  & 0.359354  & 0.995871    \\
\hline\hline
\end{tabular}
\end{center}
\label{b_table}
\end{table}




\begin{thebibliography}{99}
\bibitem{qeq1}
%
By ``correspondence with the Potts model in the limit $q \to 1$'', one
means that the thermodynamic limit is taken first for $q \ne 1$, after which
one then takes the limit $q \to 1$.  On a finite graph with $q$ set
equal to 1, the partition function is trivially given by Eq. (\ref{zq1}),
and, in particular, on a $\Delta$-regular
graph, in the limit $n(G) \to \infty$,
$f=(\Delta/2)\ln(v+1)=(\Delta/2)K$, which is analytic function of $K$
and has no non-analytic critical behavior.
\end{thebibliography}
\end{document}